\documentclass[prd,superscriptaddress,amsfonts,amssymb,amsmath,twocolumn,floatfix]{revtex4-2}
\usepackage{bm}
\usepackage{amsfonts}
\usepackage{latexsym}
\usepackage{graphicx}
\usepackage{amsmath}
\usepackage{palatino}
\usepackage{bigints}
\usepackage{mathpazo}
\usepackage{textcomp}
\usepackage{braket}
\usepackage{float}
\usepackage{booktabs}
\usepackage{dcolumn}
\usepackage{booktabs}
\usepackage{multirow}
\usepackage{hyperref}
\hypersetup{colorlinks,citecolor=blue}
\usepackage{amsmath}
\usepackage{xcolor}
\usepackage{orcidlink}
\usepackage{commath}
\usepackage{subcaption}

\def\jnl@style{\it}
\def\aaref@jnl#1{{\jnl@style#1}}

\def\aaref@jnl#1{{\jnl@style#1}}

\def\aj{\aaref@jnl{AJ}}                   
\def\apj{\aaref@jnl{ApJ}}                 
\def\apjl{\aaref@jnl{ApJ}}                
\def\apjs{\aaref@jnl{ApJS}}               
\def\apss{\aaref@jnl{Ap\&SS}}             
\def\aap{\aaref@jnl{A\&A}}                
\def\aapr{\aaref@jnl{A\&A~Rev.}}          
\def\aaps{\aaref@jnl{A\&AS}}              
\def\mnras{\aaref@jnl{Mon.~Not.~Roy.~Astron.~Soc.}}             
\def\prd{\aaref@jnl{Phys.~Rev.~D}}        
\def\prc{\aaref@jnl{Phys.~Rev.~C}}  
\def\prl{\aaref@jnl{Phys.~Rev.~Lett.}}    
\def\qjras{\aaref@jnl{QJRAS}}             
\def\skytel{\aaref@jnl{S\&T}}             
\def\ssr{\aaref@jnl{Space~Sci.~Rev.}}     
\def\zap{\aaref@jnl{ZAp}}                 
\def\nat{\aaref@jnl{Nature}}              
\def\aplett{\aaref@jnl{Astrophys.~Lett.}} 
\def\apspr{\aaref@jnl{Astrophys.~Space~Phys.~Res.}} 
\def\physrep{\aaref@jnl{Phys.~Rep.}}      
\def\physscr{\aaref@jnl{Phys.~Scr}}       
\def\commat{\aaref@jnl{Comm.~Math.~Phys.}}              
\def\science{\aaref@jnl{Science}}               
\def\cqg{\aaref@jnl{Classical Quant.~Grav.}}            
\def\jpcs{\aaref@jnl{JPCS}}                                     
\def\ijmpd{\aaref@jnl{Int.~J.~Mod.~Phys.~D}}                    
\def\grg{\aaref@jnl{Gen.~Relat.~Gravit.}}               
\def\rpp{\aaref@jnl{Rep.~Prog.~Phys.}}          
\def\npa{\aaref@jnl{Nucl.~Phys.~A}}        
\def\lrr{\aaref@jnl{Living Rev.~Rel.}}                   
\def\jcap{\aaref@jnl{J.~Cosmology Astropart.~Phys.}}    
\def\rmp{\aaref@jnl{Rev.~Mod.~Phys.}}   
\def\epjc{\aaref@jnl{Eur.~Phys.~J.~C}}

\allowdisplaybreaks[1]

\begin{document}

\title{Topological AdS black holes surrounded by Chaplygin dark fluid: from stability to geometrothermodynamic analysis}

\author{Y. Sekhmani\orcidlink{0000-0001-7448-4579}}
\email[Email: ]{sekhmaniyassine@gmail.com}

\affiliation{Ratbay Myrzakulov Eurasian International Centre for Theoretical Physics, Astana 010009, Kazakhstan.}
\author{G. G. Luciano\orcidlink{0000-0002-5129-848X}}
\email[Email: ]{giuseppegaetano.luciano@udl.cat}
\affiliation{Department of Chemistry, Physics, Environmental and Soil Sciences, Escola Polit\`ecnica Superior, Universitat de Lleida, Av. Jaume
II, 69, 25001 Lleida, Spain.}
\author{ J. Rayimbaev\orcidlink{0000-0001-9293-1838}}
\email[Email: ]{javlon@astrin.uz}
\affiliation{Institute of Fundamental and Applied Research, National Research University TIIAME, Kori Niyoziy 39, Tashkent 100000, Uzbekistan}
\affiliation{University of Tashkent for Applied Sciences, Gavhar Str. 1, Tashkent 100149, Uzbekistan }
\affiliation{ Shahrisabz State Pedagogical Institute, Shahrisabz Str. 10, Shahrisabz 181301, Uzbekistan}
\affiliation{Tashkent State Technical University, Tashkent 100095, Uzbekistan}
\author{M. K. Jasim\orcidlink{0000-0003-0888-9935}}
\email[Email: ]
{mahmoodkhalid@unizwa.edu.com}
\affiliation{Department of Mathematical and Physical Sciences, College of Arts and Sciences, University of Nizwa, Nizwa 616, Sultanate of Oman}
\author{A. Al-Badawi\orcidlink{0000-0002-3127-3453}}
\email[Email: ]{ahmadbadawi@ahu.edu.jo}
\affiliation{Department of Physics, Al-Hussein Bin Talal University, P. O. Box: 20, 71111, Ma’an,
Jordan}

\author{S.K. Maurya\orcidlink{0000-0003-0261-7234}}
  \email[Email: ]{sunil@unizwa.edu.om}
\affiliation{Department of Mathematical and Physical Sciences, College of Arts and Sciences, University of Nizwa, Nizwa 616, Sultanate of Oman}

\begin{abstract}
Implementing the concept of Dark Fluid with a Chaplygin-like equation of state within General Relativity, we construct a new higher-dimensional, static, and spherically symmetric anti-de Sitter (AdS) black hole solution. Energy conditions are explored alongside curvature singularity tools. The inspection at the level of the phase structure and $P-v$ critical behavior is carried out in the context of the extended phase space, where the cosmological constant appears as pressure. Our findings disclose non-trivial similarities between the small/large phase transition of AdS black holes surrounded by Chaplygin dark fluid and van der Waals systems' liquid/gas phase transition. This analysis offers insights into the physical interpretation of the $P-v$ diagram and identifies critical exponents that reveal the scaling behavior of thermodynamic quantities close to criticality in a universal manner. We finally deepen our understanding of the thermodynamic properties and microstructure of AdS black holes by leveraging the geometrothermodynamic formalism. Specifically, we employ tools, including Weinhold, Ruppeiner, Hendi–Panahiyan–Eslam–Momennia (HPEM), and Quevedo classes I and II. We show that each class of metrics predicts either the physical limitation point and/or the phase-transition critical points, with HPEM and Quevedo formulations providing richer information about the phase transitions. Altogether, this study contributes to advancing our knowledge of the role of Chaplygin gas in General Relativity and thoroughly examining the thermodynamic phase structure of high-dimensional AdS black holes under extreme conditions.
\end{abstract}

\maketitle

\section{Introduction}
General Relativity (GR) is currently the most theoretically comprehensive and phenomenologically appropriate framework for describing gravity. It is worth mentioning that amidst its most spell-binding predictions, gravitational waves (GWs) and black holes (BHs) represent the ultimate concrete confirmation of Einstein's theory. 
In particular, the initial detection of GWs occurred about a decade ago, originating from the fusion of a binary BH merger and the subsequent ringing of the resulting single BH~\cite{LIGOScientific:2016aoc}. Similarly, in the 1960s, detecting X-rays emitted by superheated matter orbiting a dark object provided crucial evidence for a central BH, exemplified by Cygnus X-1 within the Milky Way. However, capturing the first image of a similar spacetime oddity took nearly half a century, M87*~\cite{EventHorizonTelescope:2019dse}. 
Since then, interest in GWs and BHs has surged, offering a profound lens for exploring the cosmos at its most fundamental level. Remarkably, the study of BH physics holds promise in shedding light on the unification of GR with quantum theory and statistical mechanics~\cite{Hawking:1975vcx,Bekenstein:1973ur,Bardeen:1973gs,Hawking:1982dh}, opening up a new path to quantum gravity.

A valuable insight into the GR frame involves a particular kind of solution to Einstein's equations, namely the anti-de Sitter (AdS) BHs. On the one hand, the study of AdS BHs in the context of dual thermal field theory in an asymptotically driven manner has prompted a fluid-like description of the inherent microphysics. On the other side, the pathway for exploring the nature of equilibrium thermodynamics of BHs is a task requiring consideration of the geometric properties of BH event horizons and other relevant spacetime features~\cite{Davies:1977bgr,Cai:1998ep,Quevedo:2007mj,Quevedo:2008ry,Sahay:2010tx,Wei:2015iwa,Dehyadegari:2016nkd,KordZangeneh:2017lgs,Wei:2019yvs,Wei:2019uqg,Xu:2020gud,Ghosh:2019pwy}. 
The interplay between geometry and thermodynamics (\emph{geometrothermodynamics} - GT) is evident in the study of the BH microstructure, where the scalar curvature of the BH metric serves as a thermodynamic indicator, reflecting the behavior of interactions such as repulsion (positive curvature) or attraction (negative). The microscopic behavior of BHs has been scrutinized for a broad class of systems~\cite{Cai:1998ep,Wei:2015iwa,Wei:2019uqg,Xu:2020gud,Ghosh:2020kba} and in numerous entropic scenarios~\cite{Rani:2022xza,Jawad:2022lww,Luciano:2023fyr,Luciano23,Luciano:2023wtx,Luciano:2022hhy,LucTs,Luciano:2022ffn} within a GT scope.

Among other suggestive features of AdS BHs, 
\emph{phase transitions} and critical phenomena play a pivotal role. Pioneered in the phase space of non-rotating, uncharged Schwarzschild-AdS BHs~\cite{Hawking:1982dh}, 
discovering such phenomena paved the way for a new ground in BH thermodynamics. Within this scope, non-trivial outcomes have been reported for rotating branes~\cite{Cvetic:1999ne,Cvetic:1999rb} and charged Reissner-Nordstr\"om (RN) BHs~\cite{Chamblin:1999hg,Chamblin:1999tk}, whose first-order phase transitions exhibit critical behavior similar to a van der Waals (vdW)-type phase change (i.e. liquid-gas).

Lately, the correlation between BHs and condensed matter systems has been pursued by
identifying the cosmology constant and its conjugate quantity with the thermodynamic pressure and volume, respectively, and 
incorporating the variation of the former into the first law of BH thermodynamics for consistency with Smarr's relation (see~\cite{Caldarelli:1999xj,Kastor:2009wy,Dolan:2010ha,Dolan:2011xt,Dolan:2011jm,Cvetic:2010jb,Lu:2012xu,Kubiznak:2012wp,Kastor:2009wy} for further discussion). In the ensuing picture \emph{extended phase-space}, the mass of BH is recognized as enthalpy instead of internal energy, and diverse thermodynamic quantities (such as adiabatic compressibility, specific heat at constant pressure, etc.) could be computed using standard thermodynamic machinery~\cite{Dolan:2010ha,Dolan:2011xt,Dolan:2011jm}. Based on this approach, the equation of state (EoS) $P = P(V, T)$ has been carried out for a rotating charged AdS BH, focusing on analogies with the vdW $P-V$ diagram. The generous amount of recent work by~\cite{Kubiznak:2012wp} has brought progress in identifying first-order transitions of charged BHs with standard liquid-gas phase transitions by scrutinizing the behavior of the Gibbs free energy of an RN-AdS BH in the canonical (i.e., fixed-charge) ensemble.

Forecasting critical points in a BH system is paramount for phase transition analysis.  Toward this end, GT supplies a powerful description based on a thermodynamic metric built up from a suitable thermodynamic potential and its derivative. In this framework, Weinhold formalism provides the first attempt to assess a thermodynamic system's critical phase transition based on geometric notions~\cite{Weinhold1,Weinhold2}. On the other hand, Ruppeiner geometry stands out as an alternative tool applied to the exploration of critical thermodynamic characteristics~\cite{Rupp1,Rupp2}. Comparatively speaking, the Weinhold geometry is linked to the Ruppeiner geometry by a conformal factor expressed in terms of an inverse of temperature~\cite{WR}. Despite many virtues, both approaches are not invariant under Legendre transformation~\cite{salamon1983group,PhysRevA.41.3156}. Efforts to construct a Legendre-invariant metric in thermodynamic phase space
are due to Quevedo~\cite{Quevedo:2011np,Quevedo:2006xk,Que2} and Hendi et al.~\cite{Hendi:2015rja, Hendi:2015fya, Hendi:2015xya, Hendi:2015hoa, Hendi:2015pda}. Another perspective appears in~\cite{Mansoori:2013pna, Mansoori:2014oia, Sekhmani:2023qqe} due to Mansoori et al.

Interestingly, exploring BHs extends beyond their isolated existence, considering their interactions with the surrounding environment, including radiation and various exotic forms of matter. These interactions cause/manifest as spacetime perturbations in the form of GWs~\cite{Fabbri:2005mw,LIGOScientific:2016aoc,Grumiller:2022qhx}. One of these matter-energy sectors - the so-called \emph{dark-energy} - is believed to be responsible for the observed accelerated expansion of the current Universe \cite{SupernovaSearchTeam:1998fmf,Dodelson:2003ft}. Alongside the \emph{dark matter}, which is introduced to make up for the matter deficit in the context of the formation of galaxies, gravitational lensing and large-scale observable structure of the Universe \cite{Garrett:2010hd}, such two components account for roughly the $95\%$ of the energy budget in the Cosmos. Alternative models for Dark Sector include, for example, axions, modified gravity, or quintessence \cite{Arbey:2021gdg,Motta:2021hvl}. As a consequence, there exists a plethora of extensive studies involving BHs in such 'dark' surroundings and environments, mainly in four dimensions \cite{Gogoi:2023ntt,Sekhmani:2023ict,Gogoi:2023ntt,Gogoi:2023ffh,Kiselev:2002dx}.

In dimensions greater than four, AdS BHs and related entities play a fundamental role in the gravity/gauge duality with significant implications for holography \cite{Lunin:2001jy,Horowitz:2012nnc}. 
Additionally, alternative models such as Chaplygin gas and extensions in the form of dark fluid (CDF) offer heuristic approaches to unify dark energy and dark matter into a single entity, addressing cosmological phenomena such as the accelerated expansion, the Hubble tension and the growth of cosmic perturbations~\cite{Bean:2003ae,Carturan:2002si,Amendola:2003bz,vomMarttens:2017cuz,Sengupta:2023yxh,Abdullah:2021tee}. 
It is worth noting that while the CDF is commonly employed in cosmological studies, its EoS can be derived naturally within string theory and, in particular, from the Nambu-Goto action for a d-brane~\cite{Ogawa:2000gj,Bordemann:1993ep}. Moreover, the Chaplygin gas has been shown to have a supersymmetric generalization~\cite{Jackiw:2000cc}. Additionally, it has been suggested that the Chaplygin gas could arise due to our four-dimensional Universe being immersed in a multi-dimensional bulk~\cite{Gorini:2004by}. Consequently, it is conceivable that a Chaplygin-like dark fluid could exist as a naturally occurring substance rather than merely serving as a phenomenological model crafted for cosmological purposes. Based on these research lines, investigations have explored gravitational theories, exploring scenarios where charged static spherically-symmetric BHs interact with CDF~\cite{Li:2019lhr}. This inquiry has expanded to encompass modified Chaplygin gas models, probing the stability of BHs within different gravitational theories~\cite{Li:2022csn}. Delving into the thermodynamic aspects, efforts have been dedicated to phase transitions and critical behavior of BHs surrounded by CDF~\cite{Li:2023zfl,Aviles:2012et,Sekhmani:2023plr}, hinting at the potential for natural existence beyond theoretical conjecture.

Motivated by the intricate interplay between higher-dimensional BHs and dark sectors, 
this work aims to investigate the phase structure and critical behaviors of $d$-dimensional AdS BHs surrounded by Chaplygin-like fluids from both thermodynamic and geometric perspectives. In this sense, our analysis represents an advance compared to previous literature~\cite{Li:2023zfl}, both in terms of multidimensional generalization and geometrothermodynamic treatment, which allows for an impartial perspective on thermodynamic systems. This facilitates examining phase transition stability, patterns, and critical points. 

The manuscript is organized as follows: in Sec.~\ref{sec2}, we introduce field equations for a spherically symmetric metric in the presence of CDF structure, featured by the Chaplygin-like EoS $p=-\gamma / \rho$ \cite{Kamenshchik:2001cp} and deliberate on the primary characteristics of the resulting equations. In Sec.~\ref{ex}, after obtaining the exact solutions for the radial function $\mathcal{G}_e(r)$, we discuss the main characteristics of the resulting BH solutions, such as classical energy conditions and curvature singularities. Sec.~\ref{secther} analyzes the corresponding thermodynamics based on the first law and the Smarr relation. Specifically, we examine the relevant thermal stability and critical phenomena by studying heat capacity behavior. Sec.~\ref{sec4} takes care of studying the $P$-$v$ phase transition. Applying GT tools is useful to show a link through the space of phase transition points and takes place in Sec.~\ref{sec5}. Finally, Sec.~\ref{conc} presents the summary and conclusions of our work.

\section{Cosmological dark fluid with EoS: $p =-\gamma/\rho$}
\label{sec2}

This section is devoted to inspecting the contribution of CDF background in the context of GR. The analysis aims to explore BH solutions with a different topology in higher dimensions. For this reason, we consider the following $d$-dimensional action:
    \begin{equation}\label{action}
    \mathcal{I}=\int d^dx\sqrt{-g}\left[\frac{1}{2\kappa^2}(\mathcal{R}-2\Lambda)\right]+\mathcal{I}_{CDF},
\end{equation}
where $\mathcal{R}$ is the Ricci scalar, $g=\det (g_{\mu\nu})$ the determinant of the metric tensor $g_{\mu\nu}$, $\Lambda$ the negative cosmological constant and $\mathcal{I}_{CDF}$  the contribution sourced by CDF background. Hereafter, we assume $\kappa=8\pi G=1=c$, where $G$ and $c$ denote the Newtonian gravitational constant and speed of light, respectively. 

From then on, varying the action $(\ref{action})$ results in the following field equations:
 \begin{equation}
 \mathcal{I}_{\mu\nu}=\mathcal{R}_{\mu\nu}-\frac{1}{2}g_{\mu\nu}(\mathcal{R}-2\Lambda)-\kappa\, \mathcal{T}_{\mu\nu}=0,
 \end{equation}
where $\mathcal{T}_{\mu\nu}$ is the energy-momentum tensor for CDF. 

We consider a static, spherically and symmetric $d$-dimensional metric ansatz with $g_{tt}\,g_{rr}=-1$ containing only one unknown function, $\mathcal{G}_e(r)$
 \begin{align}
    ds^2=-\mathcal{G}_e(r)\,\mathrm{d}t^2+\mathcal{G}_e(r)^{-1}\,\mathrm{d}r^2+r^2\,\mathrm{d}\Omega_{k}^2,
 \end{align}
where $\mathrm{d}\Omega_{k}^2$ designates the line element of an $(d-2)$-dimensional hypersurface $\Sigma$ with constant curvature $(d-2)(d-3)k$. It is given by

    \begin{eqnarray}
        \mathrm{d}\Omega_{k}^2 =\mathrm{d}\theta^2+\frac{\sin^2\left(\sqrt{k}\,\theta\right)}{k}\left[\mathrm{d}\phi_1+\sum_{i =2}^{d-3}\prod_{j= 1}^{i-1}\sin^2\phi_j\mathrm{d}\phi_i^2\right],
    \end{eqnarray}
where $k = 1, 0 -1$ are consecrated to spherical, flat, and hyperbolic geometries, respectively, with $\theta \in [0, \frac{\pi}{2}]$. Notice that considering $k=1$ yields a non-trivial phase-space structure with interesting critical behaviors, while no
phase-transition state occurs for $k=0,-1$. Therefore, our next analysis shall focus on the $k=1$ scenario. Moreover, it proves convenient to define the coordinates $x_i$'s by
    \begin{eqnarray}
        x_1&=&\frac{r}{\sqrt{k}}\sin\left(\sqrt{k}\,\theta\right)\prod_{j= 1}^{n-3}\sin\phi_j,\\
        x_i&=&\frac{r}{\sqrt{k}}\sin\left(\sqrt{k}\,\theta\right)\cos\phi_{d-i-1} \\ &\times & \prod_{j= 1}^{d-i-2}\sin(\phi_j),\,\, i= 2,\cdots,d-2\nonumber\\
        x_{d-1}&=&r\cos(\sqrt{k}\,\theta).
    \end{eqnarray}

For the purposes at hand, it may be useful to look at the perfect fluid form, which is characterized by the following stress-energy tensor:
\begin{equation}
    \mathcal{T}_{\mu\nu}=\left(\rho+p\right)u_\mu u_\nu+ p g_{\mu\nu}\,.
\end{equation}
Here $\rho$ and $p$ are the energy density and isotropic pressure, respectively, as measured by an observer moving with the fluid, and $u_\mu$ is its $d$-velocity vector. At this stage, several works in the context of GR are being carried out considering static, spherically symmetric solutions with the surrounding part of the perfect fluid (dust, radiation, dark energy or ghost energy) having EoS $p = \omega \rho$ ($\omega$ being a constant) \cite{Kiselev:2002dx,Li:2014ixn,Setare:2007jw,Benaoum:2012uk,Bilic:2002chg,Chen2005:qnm,Arun:2017dm,Kubizvnak:2015bh}. 
Moreover, several indications point toward the scenario where the perfect cosmological fluid surrounding the BH  can be considered anisotropic due to the gravitational effect. Accordingly, the CDF background can be modeled as an anisotropic fluid from different perspectives; one involves a scalar field $\phi $ and a self-interacting potential $U(\phi) $, with the scalar Lagrangian $L_\phi=-\frac{1}{2}\partial_\mu\phi\partial^\mu\phi-U(\phi)$ \cite{Debnath:2004cd,Mak:2005iq,Sharif:2014yga}. Secondly, the fluid portrait of the CDF may be reshaped within a tachyonic field $T$ formulated by a Born-Infeld-type Lagrangian $L_T= -V(T)\sqrt{1 +\partial_\mu T\partial^\mu T}$, with $V(T)$ being an arbitrary real function  \cite{MagalhaesBatista:2009cus,Raposo:2018rjn}. Hence, the CDF is appropriately modeled so that its radial pressure differs from the tangential one. This is consistent with anisotropic fluids, implying a covariant form of stress-energy tensor for CDF as follows \cite{Weinhold1}
\begin{equation}\label{T1}
    \mathcal{T}_{\mu\nu}=\left(\rho+p_t\right)u_\mu u_\nu-p_t g_{\mu\nu}+\left(p_r-p_t\right)\chi_\mu \chi_\nu\,,
\end{equation}
where $p_r$ is the radial pressure in the direction of $\chi_\mu$, $p_t$ is the tangential
pressure orthogonal to $\chi_\mu$, $\chi_\mu$ is the unit spacelike vector orthogonal to the velocity $u_\mu$. Both $u_\mu$ and $\chi_\mu$ satisfy the constraint $u_\mu u^\mu=-\chi_\mu \chi^\mu=1$. 

We proceed by considering the frame comoving with the fluid so that we have $u^a=\sqrt{\mathcal{G}_e(r)}\,\delta_0^a$ and $\chi^a=1/\sqrt{\mathcal{G}_e(r)}\,\delta_1^a$. On this basis, the stress-energy tensor of Eq. (\ref{T1}) can be re-expressed as follows
\begin{equation}\label{T2}
    \mathcal{T}_\mu^\nu=-\left(\rho+p_t\right)\delta_\mu^0 \delta^\nu_0+p_t\delta_\mu^\nu+\left(p_r-p_t\right)\delta_\mu^1\delta_1^\nu\,,
\end{equation}
where the term $p_r-p_t$ is called the anisotropic factor, and once it vanishes, Eq. (\ref{T2}) is set to describe the standard isotropic background.

In what follows, we assume the matter fluid is in a state across an event horizon defined by the stress-energy expression in Eq.  (\ref{T2}). Moreover, inside the horizon, i.e., $g_{rr} <0$ and $g_{tt} >0$, the space coordinate $r$ results in behaving as the time coordinate $t$. Consequently, the energy density provides $T_r^r= p_r$, while the pressure along the spatial $t$ direction yields $T_t^t=-\rho$. From the point of view of this exchange, both the energy density and the pressure are continuous if and only if the condition $p_r=-\rho$ is satisfied. Conversely, whenever $p_r\neq-\rho$ and $\rho(r_h)\neq 0$, 
One has the pressure discontinuous at the horizon, with the solution state becoming dynamic. 

Henceforth, we shall deal only with the case where $p_r=-\rho$ \cite{Kiselev:2002dx}, in which the CDF is static and, due to certain constraints on the solution, the energy density is continuous across the horizon. To be more concrete, borrowing concepts from the anisotropic fluid leads to constraining the tangential pressure $p_t$ by taking the isotropic mean over the angles and demanding $\braket{\mathcal{T}_i^{(d)j}}=p(r) \delta_i^j$. By doing so, one can obtain
\begin{equation}\label{ptt}
   p(r) =p_t+\frac{1}{d-1}\left(p_r-p_t\right),
\end{equation}
where the conventional identity $\braket{\delta_i^1 \delta_1^j}\equiv\frac{1}{d-1}$ is taken into account. 
Compared to similar models, the higher-dimensional formulation of the tangential pressure for the quintessence matter field can be obtained from Eq. (\ref{ptt}) in the form $p_t =\frac{1}{d-2}\left((d-1)\omega+1\right)\rho$, which is compatible with the radial pressure $p_r = -\rho$.

The CDF background is characterized by a nonlinear EoS $p = -\gamma/\rho$, where $\gamma$ is a positive definite parameter. In the $p_r = -\rho$ regime, the tangential pressure of the CDF is $p_t =\frac{1}{d-2}\rho(r)-\frac{(d-1)\gamma}{(d-2)\rho(r)}$. Therefore, the stress-energy tensor component of the CDF background can be expressed, as
\begin{align}
    \mathcal{T}_t^t&=\mathcal{T}_r^r=-\rho\label{m1},\\
\mathcal{T}_{\theta_{1}}^{\theta_{1}}&=\mathcal{T}_{\theta_{i}}^{\theta_{i}}=\frac{1}{d-2}\rho(r)-\frac{(d-1)\gamma}{(d-2)\rho(r)}\label{m2}.
\end{align}
We shall see later that the anisotropy of the CDF subsides and that the EoS $p = -\gamma/\rho$ holds on the cosmological scale.

\section{Exact solutions}
\label{ex}
Because the spacetime is static and spherically symmetric, the $\mathcal{T}_t^t =\mathcal{T}_r^r$ condition is required. Hence, the components of the Einstein tensor with the negative cosmological constant are expressed by
    \begin{eqnarray}
\mathcal{I}_t^t=\mathcal{I}_r^r&=&\frac{1}{2r}(d-2)\mathcal{G}_e'(r)\\ \nonumber &+&\frac{1}{2r^2}(d-2)(d-3)(\mathcal{G}_e(r)-k)+\Lambda,\label{g1}\\
     \mathcal{I}_{\theta_i}^{\theta_i}= \mathcal{I}_{\theta_1}^{\theta_1}&=&\frac{\mathcal{G}_e''(r)}{2}+\frac{(d-3)\mathcal{G}_e'(r)}{r} \\ \nonumber &+&\frac{(\mathcal{G}_e(r)-k)(d-3)(d-4)}{2r^2}+\Lambda.\label{g2}
    \end{eqnarray}

To inspect analytically the relevant solution of the surrounding CDF background, it is necessary to combine the gravitational field equations and matter fluid equations~(\ref{m1}-\ref{m2}) and (\ref{g1}-\ref{g2}). Thereby, the first consideration leads to the following first-order differential equation for the unknown function, $\rho(r)$
\begin{equation}\label{ut}
    r\,\rho'(r) +(d-1)\,\rho(r)-\frac{\gamma (d-1) }{\rho(r)}=0\,,
\end{equation}
where the prime stands for the first derivative of the radial variable $r$. Eq. \eqref{ut} provides an exact solution for the energy density of CDF, as shown below
\begin{equation}
   \rho(r) =\sqrt{\gamma +\frac{Q^2}{r^{2(d-1)}}}.\label{ho}
\end{equation}
Here, $Q$ is a normalization factor carrying information on the intensity of the CDF fluid matter. A closer examination shows that Eq. (\ref{ho}) implements the result of the conservation law of the stress-energy tensor $\partial_\mu T^{\mu\nu}=0$. 

It is well noted that the CDF energy density is modified at certain limits. In particular, at large radial coordinates (i.e., $r^{2(d-1)}\gg Q^2/\gamma)$, it gives
\begin{equation}
     \rho(r)\sim \sqrt{\gamma}\,.
\end{equation}
This implies that the CDF looks like a positive cosmological constant in a large-scale scenario, and the closer it gets to the BH, the more densely they clump together due to gravitation. On the other hand, at small radial coordinates (i.e., $r^{2(d-1)}\ll Q^2/\gamma)$, one can obtain 
\begin{equation}
     \rho(r)\sim \frac{Q}{r^{(d-1)}}\,,
\end{equation}
which indicates that the CDF background acts like matter content whose energy density varies with $r^{d-1}$. 

On the other hand, choosing $r\rightarrow \infty$ leads to finding the properties $p_r\rightarrow-\sqrt{\gamma}$ and $p_{\theta,\theta^i} \rightarrow-\sqrt{\gamma}$, which state that the CDF background is set to be isotropic and obeys the state equation $p=-\gamma/\rho$ only at the cosmological scale. It should be noted that a cosmological fluid obeying a general EoS $p(\rho)=-\gamma/\rho$,  where the radial pressure satisfies $p_r = -\rho$ when surrounding a central BH,  still exhibits isotropic tendencies on the cosmological scale. As shown in Tab. \ref{de}, the anisotropic factor $p_r-p_t$ for CDF and quintessence matter in the Kiselev solution \cite{Kiselev:2002dx} reduces to zero at infinity. 

\begin{figure*}[t]
      	\centering{
       \includegraphics[scale=0.72]{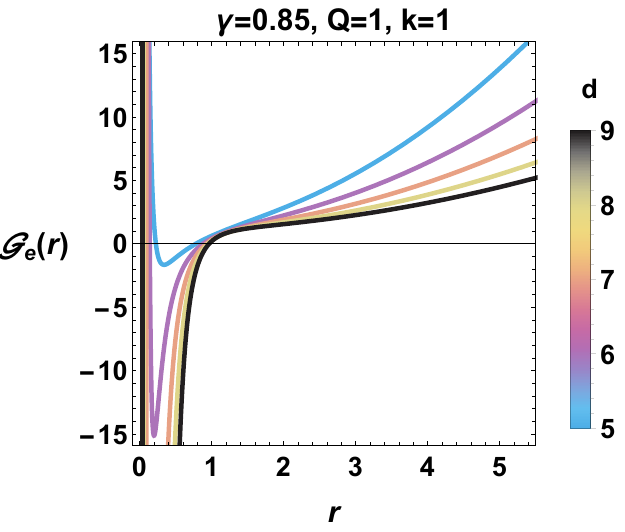} \hspace{10mm}
      	\includegraphics[scale=0.71]{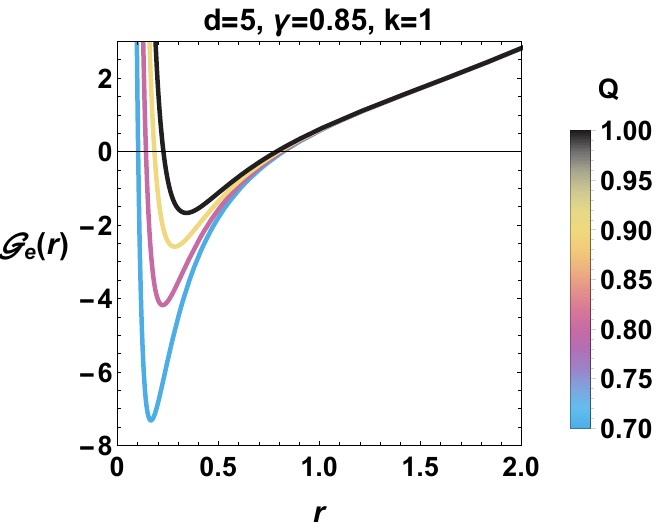} \hspace{2mm}
      }
       \centering{ 
       \includegraphics[scale=0.72]{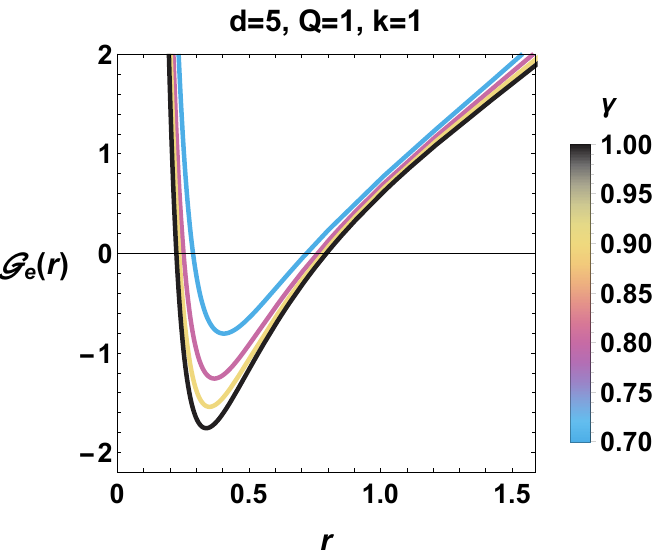}\hspace{2mm}
      \includegraphics[scale=0.7]{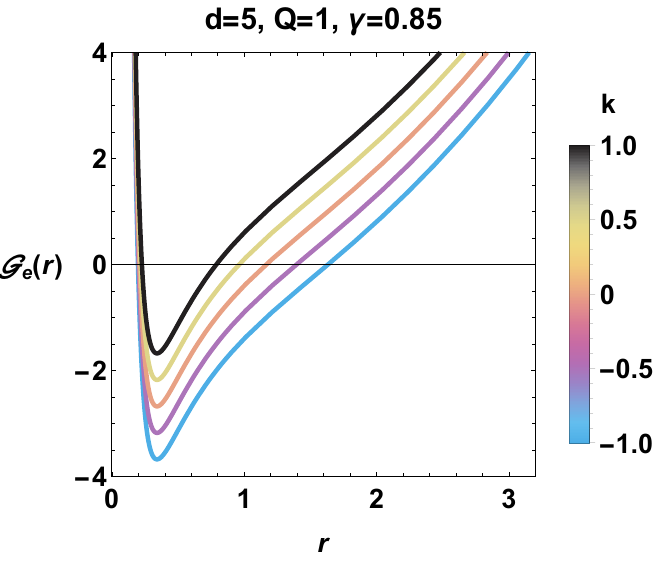} \hspace{2mm}
       }
      	\caption{Variation of the BH metric function (\ref{metric}) with respect to $r$ for various values of the parameter space with $m=1$ and $\Lambda =-4$.}
      	\label{fig1}
      \end{figure*}

We now look for the BH solution surrounded by CDF based on the considerations of the field equations~(\ref{m1}-\ref{m2}) and~(\ref{g1}-\ref{g2})  and the expression of the CDF energy density (\ref{ho}). Thus, the $(t,t)$ field equation provides the following differential equation:

    \begin{eqnarray}
    \nonumber
      (d-2) \left[(3-d) (k-\mathcal{G}_e(r))+r\, \mathcal{G}_e'(r)\right]\\ +2 r^2 \left(\sqrt{\gamma +Q^2 r^{2-2 d}}+\Lambda \right)=0\,, 
    \end{eqnarray}
which gives an analytic solution for $\mathcal{G}_e(r)$ in the following form
\begin{widetext}
    \begin{equation}\label{metric}
  \mathcal{G}_e(r) =k - \frac{m}{r^{d-3}} -\frac{2 \Lambda  r^2}{(d-2)
   (d-1)}+\frac{2 \sqrt{\gamma } Q r^{3-d} \sqrt{\frac{Q^2 r^{2-2 d}}{\gamma }+1} \text{ArcSinh}\left(\frac{Q r^{1-d}}{\sqrt{\gamma }}\right)-2 Q^2 r^{4-2 d}-2 \gamma  r^2}{(d-2) (d-1) \sqrt{\gamma +Q^2 r^{2-2 d}}}\,,
\end{equation}
\end{widetext}
where $m$ stands for an integration constant. 
      
      For a graphical description of the corresponding metric function, Fig. $\ref{fig1}$ provides the desired analysis. The present examination is, therefore, perfectly meaningful for all parameter spaces of the BH system, such as $(d, Q, \gamma, k)$. Strictly speaking, the associated roots of the metric function are analytically classified into two different types: the smaller root is related to the BH horizon. In contrast, the larger root is associated with the cosmological horizon. Interestingly, varying the parameters $d$, $Q$ and $\gamma$ has a non-trivial impact on the metric function. A closer examination shows that increasing the dimension parameter $d$ increases the distance between the two horizon radii. On the other hand, increasing the topological parameter $k$ with respect to the geometrical kind of the compact surface $ \mathrm{d}\Omega_{k}^2$ shifts the larger root toward lower values of $r$. 
      
      Concerning the dependence on $Q$ and $\gamma$, we can see that any variation of such parameters leaves the qualitative behavior of the metric function nearly unaffected.
       
Furthermore, exploring the asymptotic behavior of the metric function $\mathcal{G}_e(r)$ yields the consideration of the limit $r\rightarrow\infty$. Hence, one has
\begin{equation}
    \mathcal{G}_e(r)\rightarrow k-\frac{2r^2}{(d-2) (d-1)}\left(\Lambda +\sqrt{\gamma}\right),
\end{equation}
which shows that in the asymptotic behavior, our BH solution is tightly expressed only in terms of the cosmological constant $\Lambda$ and the CDF parameter $\gamma$. In the remainder of this paper, calculations are performed based on the AdS BH so that $\Lambda < - \sqrt{\gamma}$ can be constrained.

Next, we will investigate the characteristics of the BH solution by applying tools to predict the curvature singularities and analyze the violation or satisfaction of the Energy Conditions (EC) constraints. Moreover, we require an analysis based on scalar invariants to carry out a suitable uniqueness and singularity proof of our BH solution. These are the Ricci scalar and the Kretschmann scalar, which are defined, respectively, by
{
\begin{align}
    &\hspace{-4mm}\mathcal{R}=-\frac{2 (d-2)  \mathcal{G}_e'(r)}{r}+\frac{(d-3) (d-2) (k- \mathcal{G}_e(r))}{r^2}- \mathcal{G}_e''(r),\label{r1}\\
&\hspace{-4mm}\mathcal{R}_{\alpha\beta\mu\nu}\mathcal{R}^{\alpha\beta\mu\nu} = \frac{2 (d-2) \mathcal{G}_e'(r)^2}{r^2}+\frac{2 (d-3) (d-2) (k- \mathcal{G}_e(r))^2}{r^4}\nonumber\\
   &-\mathcal{G}_e''(r)^2.\label{r2}
\end{align}
By way of observation and because of terms $k- \mathcal{G}_e(r)$, $\mathcal{G}_e'(r)$, and $\mathcal{G}_e''(r)$ in Eqs. \eqref{r1} and \eqref{r2}, the parameter $k$ will not be present in the explicit expressions of the scalar invariants. Thus, the explicit expressions of the scalar invariants are given by  }
\begin{widetext}
\begin{align}\label{R}
    \mathcal{R}&=\frac{1}{d-2}\Biggl\{2 \Bigg(d \bigg(\frac{\gamma }{\sqrt{\gamma +Q^2 r^{2-2d}}}+\Lambda \bigg)+\frac{Q^2 r^{2-2d}}{\sqrt{\gamma +Q^2 r^{2-2d}}}\Bigg)\Biggr\}\,,
\end{align}
\end{widetext}
and
\begin{widetext}
    \begin{align}\label{RRR}
\mathcal{R}_{\alpha\beta\mu\nu}\mathcal{R}^{\alpha\beta\mu\nu}& =\frac{1}{(d-2)^2 (d-1) \left(\gamma  r^{2d}+Q^2
   r^2\right)}\Biggl\{r^{-2d} \Bigg(-4 (d-3) \left(d^2-3 d+2\right)^2 m Q^2 r^{d+3} \sqrt{\gamma +Q^2 r^{2-2 d}}\nonumber\\
   &+4 \left(d^3-3 d^2+d+3\right) Q^4 r^4+(d-3) (d-2)^4 (d-1)^2 m^2 r^2 \left(\gamma  r^{2 d}+Q^2
   r^2\right)-4 (d-3) (d-2) Q r^2\nonumber\\
   &\times\text{ArcSinh}\left(\frac{Q r^{1-d}}{\sqrt{\gamma }}\right) \bigg(\sqrt{\gamma } (d-1) r^d \sqrt{\frac{Q^2 r^{2-2d}}{\gamma }+1} \left((d-2)^2 m r^d \sqrt{\gamma +Q^2
   r^{2-2d}}-2 Q^2 r\right)\nonumber\\
   &-(d-2) Q \left(\gamma  r^{2 d}+Q^2 r^2\right) \text{ArcSinh}\left(\frac{Q r^{1-d}}{\sqrt{\gamma }}\right)\bigg)+8 \gamma  d r^{4d} \left(\gamma +\Lambda  \left(2 \sqrt{\gamma
   +Q^2 r^{2-2d}}+\Lambda \right)\right)\nonumber\\
   &+8 Q^2 r^{2 d+2} \left(2 \gamma +\Lambda  \left(d \Lambda +2 \sqrt{\gamma +Q^2 r^{2-2 d}}\right)\right)\Bigg)\Biggl\}\,.
\end{align}
\end{widetext}

Upon scrutiny of the expressions \eqref{R} and \eqref{RRR}, one can show that the BH solution described by this metric is singular. This results from any choice of the parameters $Q$, $\gamma$, and $k$ with the given constraint $d>2$. In practical terms, the presence of the singularity is due to both the mass and fluid-matter terms in the BH metric. However, contradictory scenarios can be envisaged to remove the singularity in such a way that the $d<1$ constraint is upheld, implying a nonphysical set. It is worth noting that, to carry out such a removal of singular behavior (see, for instance,~\cite{Balart:2014cga}). For the remainder of this work, we shall not think of such a situation and shall stick to the metric function \eqref{metric}. 

On the other hand, inspecting the behavior at a large distance is a useful complementary approach, given that
\begin{align}
    \lim\limits_{r\to \infty} \mathcal{R} &\approx\frac{2 d \left(\sqrt{\gamma }+\Lambda \right)}{d-2}\,,\\
    \lim\limits_{r\to \infty}\mathcal{R}_{\alpha\beta\mu\nu}\mathcal{R}^{\alpha\beta\mu\nu} &\approx\frac{8 d \left(\sqrt{\gamma }+\Lambda \right)^2}{(d-2)^2 (d-1)}\,,
\end{align}
This implies that the Ricci and Kretschmann scalar have a finite term at a large distance. To sum up, the scalar tools show that our obtained BH solution is unique, and both the AdS and the CDF backgrounds change the BH spacetime substantially.

To gain further insights into the behavior of our solution, let us focus on classical ECs, namely, the null energy condition (NEC), dominant energy condition (DEC), weak energy condition (WEC), and strong energy condition (SEC), which are defined as~\cite{Kontou:2020bta}:
 \begin{figure*}[tbh!]
    \centering
    \begin{subfigure}[b]{0.5\textwidth}
        \centering
        \includegraphics[scale=0.7]{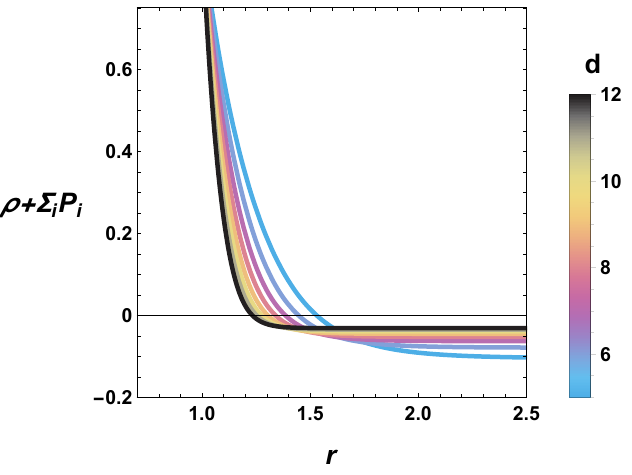}
    \end{subfigure}%
    \hfill
    \begin{subfigure}[b]{0.5\textwidth}
        \centering
        \includegraphics[scale=0.7]{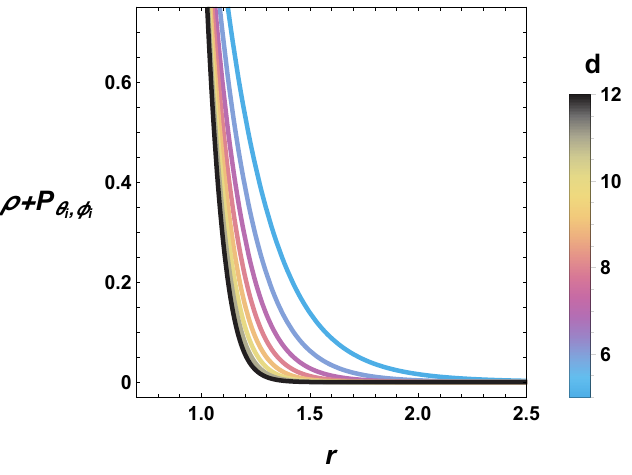}
    \end{subfigure}%
    \\
    \begin{subfigure}[b]{0.5\textwidth}
        \centering
        \includegraphics[scale=0.7]{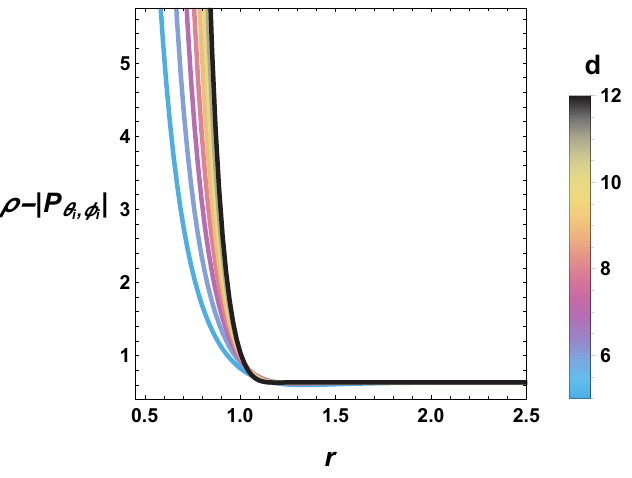}
    \end{subfigure}
    \caption{Energy conditions using $Q=1$ and $\gamma=0.2$}
    \label{fig2}
\end{figure*}
\begin{eqnarray}\label{18}
\textbf{WEC}&:& \rho\geq 0,\; \rho+P_i\geq 0,\nonumber\\[2mm]
\textbf{SEC}&:& \rho+\sum_{i}P_i\geq 0,\, \;\rho+P_i\geq 0,\nonumber \\[2mm]
\textbf{NEC}&:& \rho+P_i\geq 0,\nonumber \\[2mm]
\label{19}
\textbf{DEC}&:& \rho\geq 0, \;|P_i|\geq \rho.
\label{20}
\end{eqnarray}
Accordingly, full expressions may be furnished as follows:
\begin{align}
   \rho + P_r =0,\,\, \rho+ P_{\theta_{i},\phi_{i}}&=\frac{d-1}{d-2}\bigg(\rho-\frac{\gamma}{\rho}\bigg),\nonumber\\
   \rho+ \sum_{i}P_i&=\rho-\left(\frac{d-1}{d-2}\right)\frac{\gamma}{\rho}\,,\\
   \rho-|P_r|=0,\,\,  \rho-|P_{\theta_{i},\phi_{i}}|&=\rho-\Bigl|\frac{1}{d-2}\rho-\left(\frac{d-1}{d-2}\right)\frac{\gamma}{\rho}\Bigr|\,.\nonumber
\end{align}

  To emphasize proof of validation/satisfaction of the ECs at hand with the CDF background,
  we plot in Fig.~\ref{fig2} the behavior of $\rho+\sum_{i}P_i$, $\rho+ P_{\theta_{i},\phi_{i}}$ and $\rho-|P_{\theta_{i},\phi_{i}}|$ against the radial coordinate $r$. Careful examination shows that $\rho+ P_{\theta_{i},\phi_{i}}$ and $\rho-|P_{\theta_{i},\phi_{i}}|$ are positive definite quantities, while $\rho+\sum_{i}P_i$ changes sign at $r_0=\left(\gamma/(d-2)Q^2\right)^{\frac{1}{2(1-d)}}$, implying a positive-negative transition from small $r$ to large $r$. It is interesting to notice that the root $r_0$ is precisely the transition point for the sign of $P_{\theta_{i},\phi_{i}}$, namely, the point at which the tangential pressure changes from an attractive to a repulsive state. To sum up, the ultimate scrutiny and observation show that the CDF structure satisfies the NEC, WEC, and DEC but violates the SEC. Notice that this scenario is the same as the case of the quintessence of dark energy. From a cosmological point of view, it has been proven that any violation of the SEC in GR violates the attractive nature of gravity, as evidenced by dark energy behavior. However, this paradigm may not hold general validity in extended gravity, as shown in~\cite{Santos:2016vjg} for the case of $f(R)$ model. {Additionally, from the WEC and NEC, we infer the constraints $Q>0$, $\gamma>0$, while the DEC results in a relation between $Q$, $\gamma$, and $r$, which does not provide any meaningful information on how to fix the values of the model parameters.}
  
It is likewise useful to address the problem of ECs from a thermodynamic perspective. Notice that thermodynamic information on the BH system in the extended phase space is essentially contained in the radius $r$, which is related to the thermodynamic volume through Eq.~\eqref{r52}.
\begin{widetext}
\begin{center}
      \begin{table}[!h]    
    \begin{tabular}{lcccccc} 
    \hline\hline
       \text{Anisot. fluid}  \,&\, \text{EoS} \,&\, $p_r$ \,&\, $p_t$ \,&\, $\rho$ \,&\,\text{Asymptotic behavior} 
       \\
       \hline
       \text{Quintess. (DE)} \,&\, $p=\rho \omega$ $\left(-1<\omega<-1/3\right)$&\, $-\rho$ \,&\, $\frac{1}{d-2}\left((d-1)\omega+1\right)\rho$ \,&\, $\frac{c\,\omega(d-1)(d-2)}{4r^{(d-1)(\omega+1)}}$ \,&\, $\rho\rightarrow0$,  $p_{r,t} \rightarrow0$   \\
       \text{CDF} \,&\, $p=-\gamma/\rho$ $\left(\gamma>0\right)$ \,&\,$-\rho$ \,&\, $\frac{1}{d-2}\rho-\frac{(d-1)\gamma}{(d-2)\rho}$ \,&\, $\sqrt{\gamma+\frac{Q^2}{r^{2(d-1)}}}$ \,&\, $\rho\rightarrow\sqrt{\gamma}$,  $p_{r,t}\rightarrow-\sqrt{\gamma}$ \\
           \hline\hline
    \end{tabular}
    \caption{Quintessence and CDF in \textit{higher}-dimensional spacetime of Einstein($\Lambda$) BH.}
    \label{de}
\end{table}
\end{center}
\end{widetext}
It is interesting to compare the above considerations with recent results in the literature. For instance,  logotropic fluids like BHs may violate the SEC for high radii, as argued in~\cite{Capozziello:2022ygp}. 
    On the other hand, the study of~\cite{Fan:2016rih} reveals that, although the WEC is preserved, the SEC is always violated for regular Hayward–AdS BHs.
    In~\cite{Rodrigues:2020pem}, new solutions for regular BHs endowed with multihorizons are proposed, concluding that the SEC is never satisfied within the event horizon in all solutions. At the same time, the other ECs depend on the ratio between extreme charges of isolated solutions. A similar outcome is proposed in~\cite{Toshmatov:2017kmw} for both non-rotating and rotating  BHs in conformal gravity, where the SEC may be only satisfied for specific sizes of BHs, which depend on the new scale of the theory. 

\section{Thermodynamics}
\label{secther}
Given the gauge/gravity duality, strongly coupled gauge theories and closely allied weakly coupled string theories may be correlated. 
Thanks to the AdS/CFT correspondence \cite{Maldacena:1997re,Witten:1998qj}, wherein conformal field theory corresponds to the asymptotically AdS spacetime in a higher dimension, the thermodynamic features of a BH may be able to disclose the properties of the dual physical state. By analogy, the horizon of a BH in asymptotically AdS spacetime confers insight to the finite temperature of its dual field theory. 

Throughout this part, we will scrutinize the thermodynamic quantities of the AdS BH solution with a surrounding CDF and check the first law of thermodynamics. To determine the Hawking temperature, it is first necessary to consider the surface gravity~\cite{Kubiznak:2016qmn}, which is provided by 
\begin{equation}\label{r30}
    \chi=\left(-\frac{1}{2}\nabla_\mu\xi_\nu\nabla^\mu\xi^\nu\right)^{1/2}=\frac{1}{2}\mathcal{G}_e'\left(r_h\right),
\end{equation}
with $\xi^\mu=\partial/\partial t$ is a Killing vector for the metric. So, the formula $T=\chi/2\pi$ is the Hawking temperature, expressed in terms of the BH system parameters as
    \begin{align}\label{43}
        T&=\frac{1}{4 \pi  (d-2)}\Bigg(\frac{(d-3) (d-2) k}{r_h}-2 r_h \sqrt{\gamma +Q^2 r_h^{2-2 d}}\nonumber\\
        &-2r_h\Lambda \Bigg),
    \end{align}
where the horizon constraint $\mathcal{G}_e\left(r_h\right)=0$ has been taken into account. 

On the other hand, the entropy could stem from being computed according to the so-called area law \cite{Gibbons:1977mu,Bekenstein:1973ur}, which claims that $S_{BH}$ is equal to one-quarter of the BH's event horizon area according to
\begin{equation}\label{mass1}
   S_{BH} = \frac{\omega_{d-2}}{4}\,r_h^{d-2}\,,
\end{equation}
where $\omega_{d-2}$ represents the volume of a $(d - 2)$-dimensional unit sphere and a $(d - 2)$-dimensional hypersurface with constant negative curvature for $k = 1$ and $k = -1$,  respectively.

Next, we adopt the deductive approach of Brown and York \cite{Brown:1992br} to determine the mass. So, the metric \eqref{metric} is written in the following form:
    \begin{equation}
        \mathrm{d}s^2 =\lambda_{ab}\mathrm{d}x^a\mathrm{d}x^b= -V(r)\mathrm{d}t^2+\frac{\mathrm{d}r^2}{V(r)}+r^2\mathrm{d}\Omega_{k,d-2}^2,
    \end{equation}
with the set of the metric background as
\begin{equation}
  V_0(r)=  k  -\frac{2 \Lambda  r^2}{(d-2)
   (d-1)},\,\,\,\,\,\, d\neq5\,,
\end{equation}
where $V_0(r)$ stands for an optional function specifying the zero of the energy to prevent infinities of the mass. As long as we customize $\sigma_{ab}$ as the metric of the spatial surface $\Sigma$ in $\partial M$, and $n^a$ and $\xi^b$ as the unit normal and the timelike killing vectors of this boundary, respectively, the mass of said BH is reckoned by 
    \begin{equation}
         M=\frac{1}{8\pi}\int_\Sigma d^{d-2}\sqrt{\sigma}\biggl\{\left(K_{ab}-K\lambda_{ab}\right) -\left(K_{ab}^0-K^0\lambda_{ab}^0\right)\biggr\}n^a\xi^b\,,
         \\
    \end{equation}
with $\sigma$ being the determinant of the $\sigma_{ab}$ metric, $K^0_{ab}$ the extrinsic curvature tensor of the background metric and $\lambda^0_{ab}$ the tensor background metric. In the $r \rightarrow \infty$ limit, the mass of the AdS BH solution with a surrounding CDF is given by
\begin{equation}\label{ms}
    M=\frac{(d-2) \omega_{d-2}}{16\pi}m\,,
\end{equation}
where $m$ is identified simply by the equation $\mathcal{G}(r_h) = 0$ in Eq. \eqref{metric}. So, in terms of the parameter BH spacetime, the Arnowitt-Deser-Misner
(ADM) mass is given by
\begin{widetext}
    \begin{equation}
  M=\frac{(d-2)\,\omega_{d-2}\,r_h^{d-3}}{16\pi}    \Bigg\{k-\frac{2 r_h^2 }{(d-2) (d-1)
   }\Bigg(\frac{\gamma -\sqrt{\gamma } Q r^{1-d} \sqrt{\frac{Q^2 r^{2-2d}}{\gamma }+1} \text{ArcSinh}\left(\frac{Q r^{1-d}}{\sqrt{\gamma }}\right)+Q^2 r_h^{2-2d}}{\sqrt{\gamma +Q^2 r_h^{2-2d}}}+\Lambda\Bigg)\Bigg\}.\label{mass}
\end{equation}
\end{widetext}

A useful insight is the implementation of the first law of BH thermodynamics to achieve a proper derivation of the relevant thermodynamic quantities. Concretely, the first law can be written as~\cite{Cai:1998vy}
\begin{equation}\label{r32}
    \mathrm{d} M=T\mathrm{d}S+\sum_i \mu_i \,\mathrm{d}\mathcal{N}_i\,,
\end{equation}
where $\mu_i$ are the chemical potentials corresponding to the conserved charges $\mathcal{N}_i$. By considering the extended phase space, several crucial thermodynamic processes become relevant. Within this framework, the phenomenon of $P-V$ criticality stands out, particularly notable in its analogy to ``black hole chemistry'', where the negative cosmological constant mimics a pressure. With this aim in mind, the law of thermodynamics is as follows~\cite{Dolan:2010ha, Brown:1987dd}:
\begin{equation}\label{r40}
\mathrm{d}H=\mathrm{d}M=T\mathrm{d}S+V\mathrm{d}P+\Phi \mathrm{d}Q\,,
\end{equation}
where { $\Phi$ is the thermodynamic potential conjugated to the normalization factor $Q$ of the CDF structure and is brought in to render the first law consistent with Smarr's relation. On the other hand,} the ADM mass resembles enthalpy and the pressure $P$ is defined as~\cite{Kastor:2009wy, Gunasekaran:2012dq}
\begin{equation}\label{pl}
P=-\frac{\Lambda}{8\pi} =\frac{(d-1) (d-2)}{16\pi\ell^2}\,.
\end{equation} 
Notice that, for $d=4$, this correctly reduces to the standard relation $P=3/(8\pi \ell^2)$~\cite{Kubiznak:2012wp}.

Based on the specific volume $v =\frac{4r_h}{d-2}$ and making use of equation \eqref{ms}, the conjugate quantity of $P$ results in the following amount:
\begin{equation}
 V=\frac{\omega_{d-2}}{d-1}r_h^{d-1}\,.
\end{equation}
For the sake of the description of thermodynamic phase space quantities, the parameter function $\mathcal{G}_e(r_h, M, P, Q)$ turns to vanish for any transformation of the parameter BH system. Related remarks on this issue raise similar grounds for the constraints $\mathcal{G}_e(r_h, M, P, Q)= 0$ and $\delta \mathcal{G}_e(r_h, M, P, Q)= 0$ upon the evolution along the parameter space. On the other hand, another possibility might be considered, incorporating the mass parameter $M$ as a function of the parameters $M(r_h,\ell, Q)$ too.

The relevant thermodynamic parameters within this study are three, namely, $S$, $P$, and $Q$. It is thus straightforward to redefine $M=M(S, P, Q)$ as a complete explicit parameterised function
\begin{equation}\label{r41}
\mathrm{d}M=\bigg(\frac{\partial M}{\partial S}\bigg)_{P,Q}\mathrm{d}S+\bigg(\frac{\partial M}{\partial P}\bigg)_{S,Q}\mathrm{d}P+\bigg(\frac{\partial M}{\partial Q}\bigg)_{S,P}\mathrm{d}Q.
\end{equation}
Analogously, this is akin to a differential 1-form in parameter space $\left(S, P, Q\right)$. Consequently, all the components $\left(\partial M/\partial x^i\right)_{x^k}$ are nothing more than thermodynamic quantities expressed in the context of the extended phase space along the basis 1-form: $\lbrace \mathrm{d}S, \mathrm{d}P, \mathrm{d}Q\rbrace$. Following up this way, one can infer the following quantities: 
\begin{equation}\label{r42}
T=\bigg(\frac{\partial M}{\partial S}\bigg)_{P,Q},\,
V=\bigg(\frac{\partial M}{\partial P}\bigg)_{S,Q},\,
\phi=\bigg(\frac{\partial M}{\partial Q}\bigg)_{S,P}.
\end{equation}

Alternatively, these findings could be derived similarly by setting the following condition on the variation of $\mathcal{G}_e(r_h, M, P, Q)$ in the parameter space
\begin{eqnarray}
   0 &=&\frac{\partial \mathcal{G}_e}{\partial r_h}\mathrm{d}r_h+\frac{\partial \mathcal{G}_e}{\partial M}\mathrm{d}M+\frac{\partial \mathcal{G}_e}{\partial P}\mathrm{d}P+\frac{\partial \mathcal{G}_e}{\partial Q}\mathrm{d}Q\nonumber\\[2mm]
    &=&\mathrm{d}\mathcal{G}_e(r_h, M, P, Q)\,,
\end{eqnarray}
\label{r43}
yielding
\begin{eqnarray}\label{mr}
   \mathrm{d}M&=&\left(\frac{1}{4\pi}\frac{\partial \mathcal{G}_e}{\partial r_h}\right)\left(-\frac{1}{4\pi}\frac{\partial \mathcal{G}_e}{\partial M}\right)^{-1}\mathrm{d}r_h\nonumber\\
   &+&\left(-\frac{\partial \mathcal{G}_e}{\partial M}\right)^{-1}\left(\frac{\partial \mathcal{G}_e}{\partial P}\right)\mathrm{d}P\nonumber\\
   &+&\left(-\frac{\partial \mathcal{G}_e}{\partial M}\right)^{-1}\left(\frac{\partial \mathcal{G}_e}{\partial Q}\right)\mathrm{d}Q \,,
\end{eqnarray}
which must be in conformity with Eq.~(\ref{r40}). 
The relation~(\ref{mr}) embraces the appearance of temperature, which is geometrically defined as
\begin{equation}\label{r45}
T=\frac{1}{4\pi}\frac{\partial \mathcal{G}_e}{\partial r_h}.
\end{equation}
In turn, by consistency with the first law of thermodynamics, this provides
\begin{eqnarray}\label{r46}
\mathrm{d}S=\bigg(-\frac{1}{4\pi}\frac{\partial \mathcal{G}_e}{\partial M}\bigg)^{-1}\,\mathrm{d}r_h.
\end{eqnarray}
It should be pointed out that this expression can also be derived using Wald's formalism.  Basically, $\delta S = \delta \int\frac{\partial L}{\partial R}$ as long as $\mathrm{d}\mathcal{G}_e = 0$ is satisfied.

Furthermore, the thermodynamic volume and the conjugate potential are defined by the following formula:
\begin{align}\label{r47}
V&=\bigg(\frac{\partial M}{\partial P}\bigg)_{S,Q}=\bigg(-\frac{\partial \mathcal{G}_e}{\partial M}\bigg)^{-1}\bigg(\frac{\partial \mathcal{G}_e}{\partial P}\bigg),\\
\phi&=\bigg(\frac{\partial M}{\partial Q}\bigg)_{S,P}=\bigg(-\frac{\partial \mathcal{G}_e}{\partial M}\bigg)^{-1}\bigg(\frac{\partial \mathcal{G}_e}{\partial Q}\bigg),
\end{align}

In the case of the BH system, the enthalpy is defined by the system's total mass. Consequently, in terms of the parameters of the BH system and the context of the extended phase space, the thermodynamic volume and the { potential conjugated to the normalization factor $Q$ of the CDF structure} can be formulated as
\begin{align}\label{r52}
V&=\left(\frac{\partial M}{\partial P}\right)_{S,Q}=\frac{\omega_{d-2}}{d-1}r_h^{d-1}\,,
\\[2mm]
\phi&=\left(\frac{\partial M}{\partial Q}\right)_{S,P}=\frac{\omega_{d-2}  \sinh ^{-1}\left(\frac{Q r_h^{1-d}}{\sqrt{\gamma }}\right)}{8 \pi 
   (d-1)}\,,
\end{align}
while the Hawking temperature is given by
    \begin{equation}\label{tex}
        T=\frac{\frac{(d-3) (d-2) k}{r_h}-2 r_h \left(\sqrt{\gamma +Q^2 r_h^{2-2 d}}-8 \pi  P\right)}{4 \pi  (d-2)}\,.
    \end{equation}
{According to Euler's theorem~\cite{Kastor:2009wy,Altamirano:2014tva}, the Smarr formula can be constructed for the CDF structure in the framework of GR by considering the dimensional analysis, $[M]=(d-3)$, $[\Lambda]=-(d-2)$, $[S]=(d-2)$, $[Q]=(d-3)$ and $[\gamma]=-4$. We find} 
    \begin{equation}\label{r48}
(d-3) M=(d-2) TS-2PV+(d-3)\phi Q\,,
\end{equation}
which gives another combination among thermodynamic quantities. Notably, the above relation correctly reproduces the corresponding condition of~\cite{Li:2023zfl} for a $(3+1)$ dimensional BH.

\subsection{Thermal stability}
\label{sec3}
As part of the canonical ensemble, heat capacity represents an important thermodynamic quantity, providing information on the thermal state of BHs. Heat capacity implies three specific and tantalizing items of evidence. First, this quantity's discontinuous nature hints at the possible thermal phase transitions that the system may undergo. Second, a further feature concerns the sign of the heat capacity. The sign indicates whether the system is thermally "stable" or "unstable." Positivity, in other words, generates thermal stability, while the opposite indicates instability~\cite{Sahabandu:2005ma, Cai:2003kt}. The third point is that the potential roots offer valuable clues indicating sign changes, which could signify stable or unstable states or boundary points. 

The next analysis is devoted to computing these premises' relevant heat capacity in the extended phase space. Toward this end, we remind that heat capacity (at constant $P$ and $Q$) can be calculated as~\cite{Kubiznak:2016qmn}:
\begin{equation}\label{r34}
C_{P,Q} =\frac{\left(\frac{\partial M}{\partial S}\right)_{P,Q}}{\left(\frac{\partial^2 M}{\partial S^2}\right)_{P,Q}} =T\left(\frac{\partial T}{\partial S}\right)_{P,Q}^{-1}\,,
\end{equation}
which gives in the present case
 \begin{figure*}[t!]
    \centering
    \begin{subfigure}[b]{0.5\textwidth}
        \centering
        \includegraphics[scale=0.7]{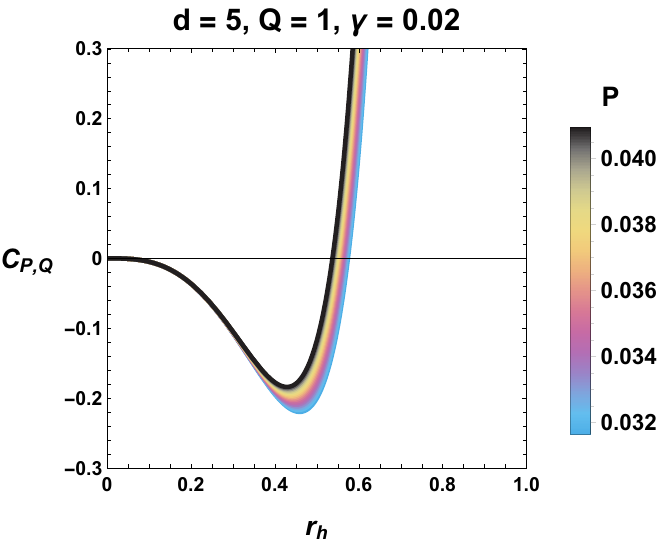}
        \caption{Physical limitation point}
        \label{ch1}
    \end{subfigure}%
    \hfill
    \begin{subfigure}[b]{0.5\textwidth}
        \centering
        \includegraphics[scale=0.74]{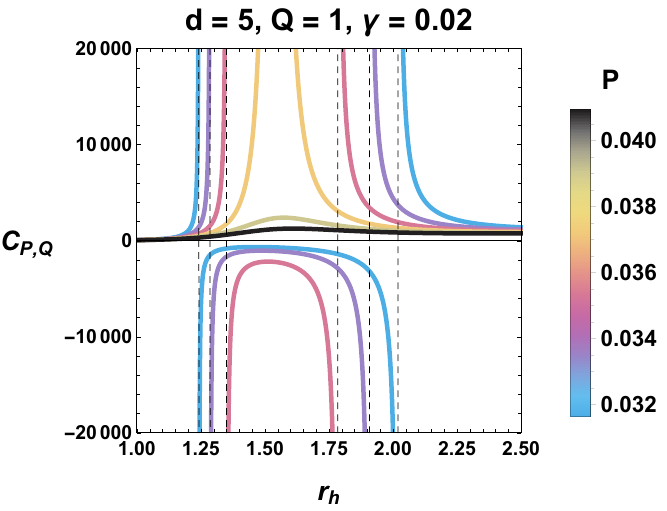}
        \caption{phase transition critical point}
        \label{ch2}
    \end{subfigure}%
    \\
      \begin{subfigure}[b]{0.5\textwidth}
        \centering
         \includegraphics[scale=0.7]{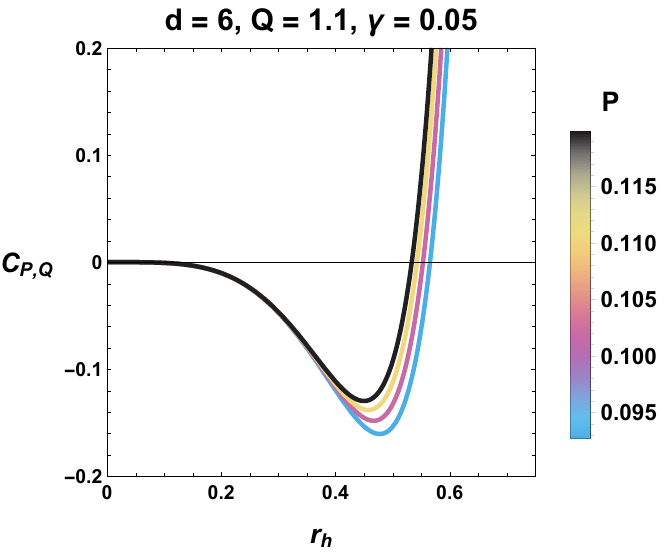}
        \caption{Physical limitation point}
        \label{ch3}
    \end{subfigure}%
    \hfill
    \begin{subfigure}[b]{0.5\textwidth}
        \centering
        \includegraphics[scale=0.74]{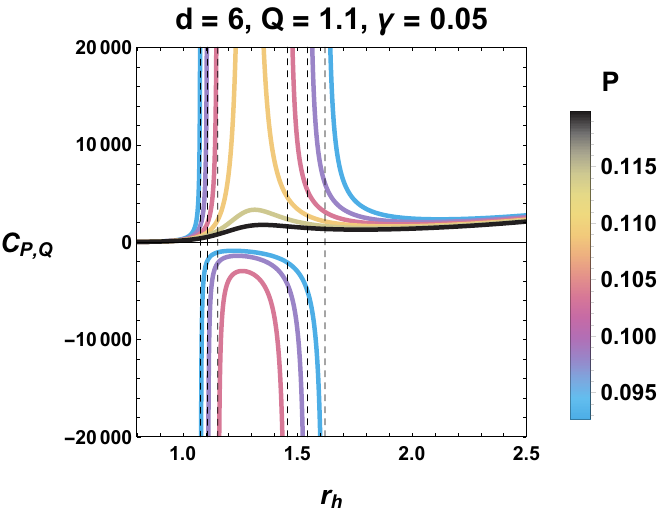}
        \caption{phase transition critical point}
        \label{ch4}
    \end{subfigure}%
    \caption{Variation of heat capacity $C_{P, Q} $ (\ref{ch}) as a function of $r_h$.} 
    \label{fig3}
\end{figure*}
\begin{widetext}
     \begin{equation}\label{ch}
            C_{P,Q}=\frac{(d-2) \omega_{d-2}  \left(r_h^{2 d} \left((3-d) (d-2) k \sqrt{\gamma +Q^2 r_h^{2-2 d}}+2 r_h^2 \left(\gamma -8 \pi  P \sqrt{\gamma +Q^2 r^{2-2 d}}\right)\right)+2 Q^2 r_h^4\right)}{4 r_h^{d+2} \left((d-3) (d-2)
   k \sqrt{\gamma +Q^2 r_h^{2-2 d}}+2 r_h^2 \left(\gamma -8 \pi  P \sqrt{\gamma +Q^2 r_h^{2-2 d}}\right)\right)-8 (d-2) Q^2 r_h^{6-d}}.
        \end{equation} 
\end{widetext}

As argued above, in the subject of BH physics, the corresponding roots of the heat capacity $(C_{P,Q} = T = 0)$ indicate a separating region between physical $(T > 0)$ and non-physical $(T < 0)$ systems. These roots are commonly referred to as ``physical limitation points''. On the other hand, the set of heat capacity divergence points represents the critical phase transition points of the BHs~\cite{Tzikas:2018cvs}. In this way, the critical phase transition points and limitation points of the BHs are explicitly designed, taking into account the following constraints:
  \begin{itemize}
    \item $T=\left(\frac{\partial M}{\partial S}\right)_{P,Q}=0$ \hspace{0.4cm} \text{physical limitation points}\, \\
    \vspace*{2mm}
   \item    $\left(\frac{\partial^2 M}{\partial S^2}\right)_{P,Q}=0$\hspace{0.3cm}\text{second order phase transition}\,.
    \end{itemize}
    
To find the physical limitation and critical phase transition points, we consider the above relations alongside the definition~\eqref{mass} and~\eqref{mass1} with Eq.~\eqref{pl}. After some algebra, we get
\begin{widetext}
    \begin{align}
   &\left(\frac{\partial M}{\partial S}\right)_{P,Q}=\frac{1}{4 \pi  (d-2)}\Bigg(\frac{(d-3) (d-2) k}{r_h}-2 r_h \left(\sqrt{\gamma +Q^2 r_h^{2-2 d}}-8 \pi  P\right)\Bigg)=0,\\
   &\left(\frac{\partial^2 M}{\partial S^2}\right)_{P,Q}  =\frac{2 (d-2) Q^2 r_h^{5-3 d}-r_h^{1-d} \left((d-3) (d-2) k \sqrt{\gamma +Q^2 r_h^{2-2 d}}+2 r_h^2 \left(\gamma -8 \pi  P \sqrt{\gamma +Q^2 r_h^{2-2 d}}\right)\right)}{\pi  (d-2)^2 \omega_{d-2}  \sqrt{\gamma +Q^2
   r_h^{2-2 d}}}=0\,.
\end{align}
\end{widetext}

The analytic resolution of these equations 
is hindered by various technicalities. Thus, we apply the numerical approach to find the physical limitation points and phase transition points corresponding to a given parameter space. Numerical solutions are shown in Tabs. $\ref{Tab1}$- $\ref{Tab2}$ for two given parameter spaces. 
\begin{widetext}
\begin{center}
      \begin{table}[!h]      
    \begin{tabular}{lccccccc} 
    \hline\hline
       $d$  \,&\, $Q$ \,&\, $\gamma$ \,&\,  $P$ \,&\, $r_1^\star$ \,&\,  $r_2^\star$ \,&\,  \text{Number of points}
       \\
       \hline
       5 \,&\, 1 \,&\, 0.02 \,&\,   0.0316415$<P_c$ \,&\,0.555145 \,&\,$\emptyset$ \,&\, 1\\
       5 \,&\, 1 \,&\, 0.02 \,&\,    0.0372253$=P_c$ \,&\,0.551774 \,&\, $\emptyset$ \,&\, 1\\
        5 \,&\, 1 \,&\, 0.02 \,&\,   0.0409478 $>P_c$\,&\,0.549601\,&\, $\emptyset$\,&\, 1
        \\
         6 \,&\, 1.1 \,&\, 0.05 \,&\,    0.0926834$<P_c$ \,&\,0.547618 \,&\,$\emptyset$ \,&\, 1\\
       6 \,&\, 1.1 \,&\, 0.05 \,&\,   0.109039$=P_c$ \,&\,0.544544 \,&\, $\emptyset$ \,&\, 1\\
        6 \,&\, 1.1 \,&\, 0.05 \,&\,    0.119943 $>P_c$\,&\,0.542566\,&\, $\emptyset$\,&\, 1
        \\
     \\
           \hline\hline
    \end{tabular}
    \caption{Physical limitation points with $k=1$.}
    \label{Tab1}
\end{table}
\end{center}
\end{widetext}

To gain an insight into the behavior of the heat capacity at constant pressure, we plot in Fig.~$\ref{fig3}$ the behavior of $C_{P,Q}$ as a function of the horizon radius $r_h$. Our analysis covers two choices in the parameter space, as represented by Figs. $\ref{ch1}$-$\ref{ch2}$ and Figs. $\ref{ch3}$-$\ref{ch4}$, respectively. 
Moreover, Tab.~$\ref{Tab1}$- $\ref{Tab2}$ includes data on physical limitations and divergent points. 
From Figs. $\ref{ch1}$-$\ref{ch3}$, we notice that, whatever the value of the pressure $P$, 
we always have one root for the finite value of $r_h$, corresponding to a physical limitation point. Thus, inspecting the thermal local stability gives two physical states: unstable for $r_h<r_1^\star$ and stable above.

On the other hand, the occurrence of physical phase transition points appears clear in the parameter spaces in Figs.~$\ref{ch2}$-$\ref{ch4}$.  For $P$ below a certain threshold $P_c$ (\emph{critical pressure}, see the definition in Eq.~\eqref{PC} below), the heat capacity has two divergent points. Therefore, the set of these points can generate three intervals in the space of the horizon radius $r_h$. In this way, BHs are locally thermal stable for regions $r_h < r_1^\text{div}$ (small BH region) and $r_h>r_2^\text{div}$ (large BH region), while thermal instability occurs for $r_1^\text{div} < r_h<r_2^\text{div}$ (intermediate BH region). As stressed above, this heat capacity behavior indicates a phase transition between the small and large BH phases.
Furthermore, as the pressure increases and approaches the critical value $P_c$, the number of divergent points reduces to a single divergent point (\emph{critical point}, see Eqs.~\eqref{c1}-\eqref{c2} below). In this case, the heat capacity is always positive and the intermediate BH phase degenerates in a single point. Finally, for $(P>P_c)$, the critical behavior is no longer present due to the absence of such a discontinuity provided by the divergent point. This situation keeps the BHs thermally stable. In a nutshell, our BH solution definitely remains in a thermally stable state. More quantitative studies on the critical behavior of BHs will be provided in the next section. 
      \begin{widetext}
\begin{center}
      \begin{table}[ht]     
    \begin{tabular}{lccccccc} 
    \hline\hline
       $d$  \,&\, $Q$ \,&\, $\gamma$ \,&\,  $P$ \,&\, $r_1^{\text{div}} $ \,&\,  $r_2^{\text{div}} $ \,&\,  \text{Number of points}
       \\
       \hline
       5 \,&\, 1 \,&\, 0.02 \,&\,   0.0316415$<P_c$ \,&\,1.2422 \,&\,2.01935 \,&\, 2\\
       5 \,&\, 1 \,&\, 0.02 \,&\,    0.0372253$=P_c$ \,&\,1.54081 \,&\, $\emptyset$ \,&\, 1\\
        5 \,&\, 1 \,&\, 0.02 \,&\,   0.0409478 $>P_c$\,&\, $\emptyset$\,&\, $\emptyset$\,&\, 0
        \\
         6 \,&\, 1.1 \,&\, 0.05 \,&\,    0.0926834$<P_c$ \,&\,1.07819 \,&\,1.6213 \,&\, 2\\
       6 \,&\, 1.1 \,&\, 0.05 \,&\,   0.109039$=P_c$ \,&\,1.28493 \,&\, $\emptyset$ \,&\, 1\\
        6 \,&\, 1.1 \,&\, 0.05 \,&\,    0.119943 $>P_c$\,&\,$\emptyset$\,&\, $\emptyset$\,&\, 0\\
      \\
           \hline\hline
    \end{tabular}
    \caption{Phase transition points with $k=1$.}
     \label{Tab2}
\end{table}
\end{center}
\end{widetext}
\section{$P-V$ criticality}
\label{sec4}
This section studies $P-V$-criticalities for AdS BHs surrounded by a CDF background. For that purpose, using the expression for the Hawking temperature in the realm of the BH chemistry concept \eqref{tex}, we may obtain the equation of state in the following form:
\begin{widetext}
\begin{align}\label{pt}
P&= -\frac{k}{8 \pi  r^2}\left(3+\frac{d}{2}\left(d-5\right)\right)+\frac{T}{4r_h}\left(d-2\right)+\frac{\sqrt{\gamma +Q^2 r^{2-2 d}}}{8 \pi },
\end{align}
\end{widetext}
\begin{figure*}[tbh!]
      	\centering{
       \includegraphics[scale=0.75]{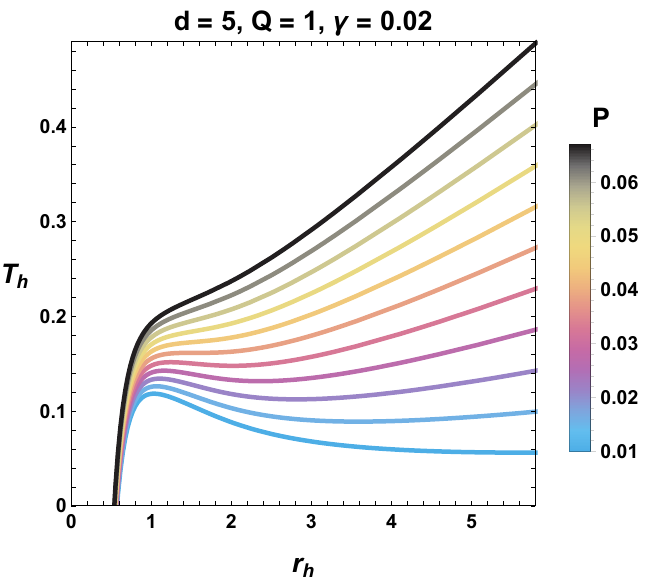} \hspace{2mm}
      	\includegraphics[scale=0.75]{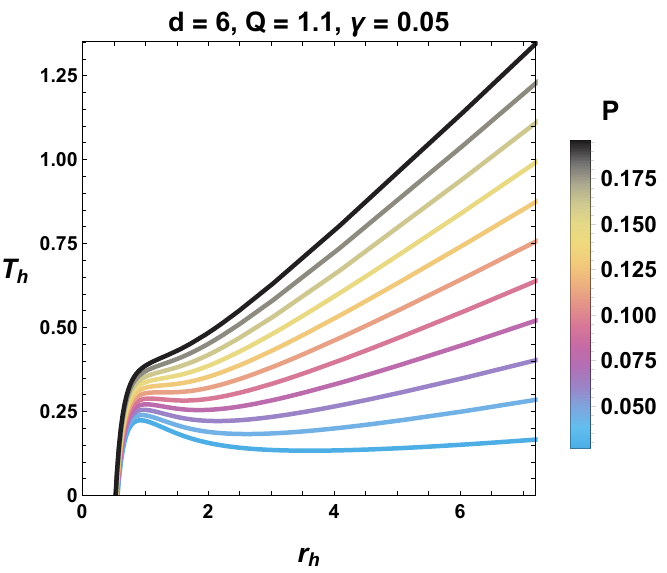} \hspace{2mm}
      }       
      	\caption{Isobaric curve $T_h$-$r_h$ diagrams of the BH system for given parameter space values.}
      	\label{fig5}
      \end{figure*}
which obviously shows the non-trivial impact of parameters $d$, $k$, $\gamma$, and $Q$ on thermodynamic behavior. As intended, we can define the specific volume $v =\frac{4r_h}{d-2}$, wherewith the pressure is cast in the standard form $P = \frac{1}{v}\, T + \mathcal{O}(v)$. Moreover, since the thermodynamic volume $V \propto r^{d-1}_h$, the critical point associated with phase transitions can be inspected considering the following constraints:
\begin{equation}\label{c1}
\bigg(\frac{\partial P}{\partial r_h}\bigg)_T=0,\quad \bigg(\frac{\partial^2 P}{\partial r_h^2}\bigg)_T=0\,,
\end{equation}
or alternatively,
\begin{equation}\label{c2}
\bigg(\frac{\partial T}{\partial r_h}\bigg)_P=0,\quad \bigg(\frac{\partial^2 T}{\partial r_h^2}\bigg)_P=0.
\end{equation}

Implementing Eqs.~\eqref{c1}-\eqref{c2} and exploring the characteristics of the critical phase transition can lead to the set of higher-dimensional critical triplets $(T_c, P_c, r_c)$ in the guise of
\begin{figure*}[tbh!]
      	\centering{
       \includegraphics[scale=0.8]{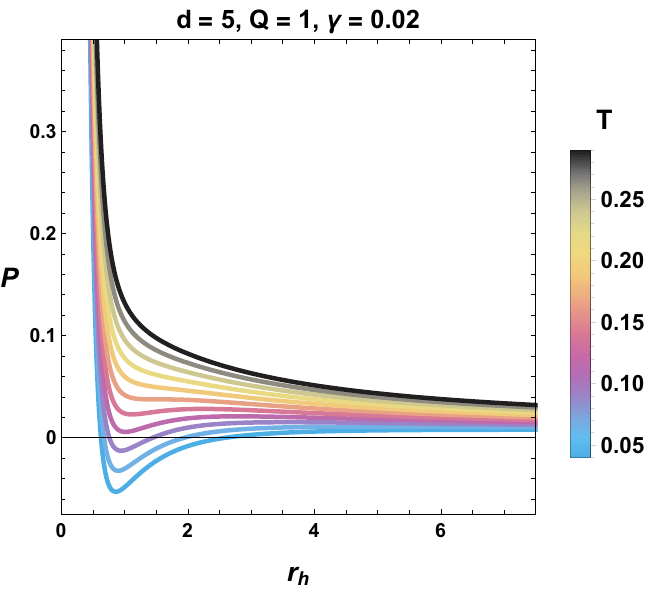} \hspace{2mm}
      	\includegraphics[scale=0.77]{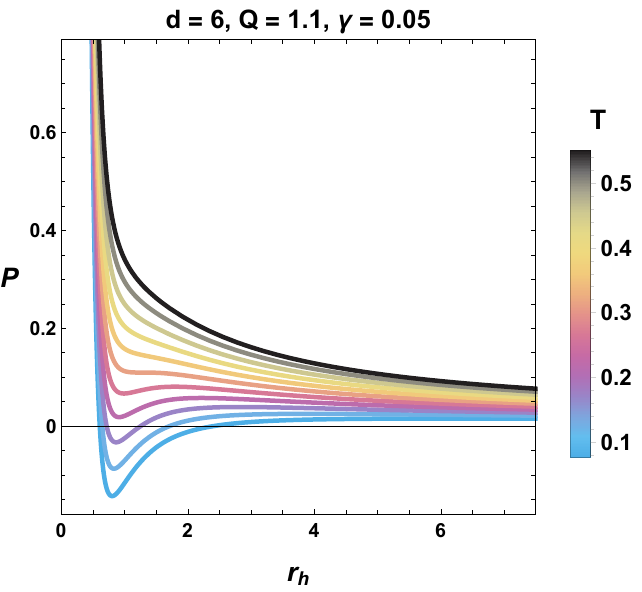} \hspace{2mm}
      }
       
      	\caption{Isotherme curve $P$-$r_h$ diagrams of the BH system for given parameter space values. }
      	\label{fig6}
      \end{figure*}
      \begin{widetext}
          \begin{align}
        T_c&= \frac{1}{2 \pi  (d-2) r_c}\bigg((d-3) (d-2) k-\frac{(d-1) Q^2 r_c^{4-2 d}}{\sqrt{\gamma +Q^2 r_c^{2-2 d}}}\bigg),
    \\
        \label{PC}
        P_c&=  \frac{1}{16 \pi  \sqrt{\gamma +Q^2 r_c^{2-2 d}}}\Bigg(r_c^{-2 (d+1)} \bigg(r_c^{2 d} \left((d-3) (d-2) k \sqrt{\gamma +Q^2 r_c^{2-2 d}}+2 \gamma  r_c^2\right)-2 (d-2) Q^2 r_c^4\bigg)\Bigg),
      \end{align}
      and
\begin{align}\label{cc}
\frac{1}{\left(\gamma  r_c^{2 d}+Q^2 r_c^2\right)^2}\Bigg((d-1) Q^2 \sqrt{\gamma +Q^2 r_c^{2-2 d}} \bigg((d-2) Q^2 r_c^2+\gamma  (2 d-3) r_c^{2 d}\bigg)\Bigg)-\frac{(d-3) (d-2) k}{r_c^4}=0.
\end{align}
\end{widetext} 

As discussed above, solving  Eq. \eqref{cc} analytically is tough. However,  valuable insights can still be gained by applying the numerical approach.  
In Tab. $\ref{Tab3}$, we collect the numerical sets of critical points for multiple valued parameter spaces. It can be seen that critical pressure and temperature increase in functions of spacetime dimensions while critical horizon radius decreases. The universal ratio $\frac{P_c\,r_c}{T_c}$ is an increasing function of spacetime dimensions, the same as RN-AdS BHs in Einstein gravity \cite{Kastor:2009wy,Kubiznak:2012wp,Cvetic:2010jb}. Similar observations can be made for the variation of the parameters $Q$ and $\gamma$ given critical pressure, temperature, critical horizon radius, and the universal ratio $\frac{P_c\,r_c}{T_c}$, respectively. 
     
     To emphasize the core characteristics of $P$-$V$ criticality, the data sets $T$-$r_h$ and $P$-$r_h$ are particularly handy. Fig. $\ref{fig5}$ shows the isobaric curve along the $T-r_h$ diagram for a definite valued spectrum of the pressure $P$. In line with the previous discussion at the end of Sec.~\ref{secther}, the case $P<P_c$ entails two extreme points (a local maximum and minimum, respectively) generating three branches, i.e., the small BH branch, the intermediate BH branch, and the large BH branch. In particular, the small and large branches of the BH are characterized by a positive slope, which implies the system's positive energy capacity and thermal stability.  On the other hand, the intermediate branch has a negative slope, i.e., the BH is thermally unstable due to the negativity of the heat capacity.
     By contrast, as $P$ tends to the critical value $P_c$, the two extreme points collapse into a single inflection point, while the $P>P_c$ exhibits no extreme point. 
     
     Further, Fig.~$\ref{fig6}$ shows the critical behavior $P$-$r_h$ via an isothermal curve process through a temperature spectrum $T$. Interestingly enough, the diagram is very similar to the VdW liquid-gas system. In particular, the $T<T_c$ scenario predicts the occurrence of a small-large BH phase transition akin to the VdW liquid-gas phase transition. 
     At the critical temperature $(T=T_c)$, the number of phase transitions is confined to one, i.e., the first-order phase transition blends with the second-order phase transition analogous to a real gas system. However, until $(T>T_c)$, the system reveals a single-phase behavior and is effectively an ideal gas with an emptiness of phase transitions.

  \begin{table*}[ht!]
    \begin{tabular}{lcccccccccc} 
    \hline\hline
        $d$  \,& $Q$ \,& $\gamma$ \,&  $r_c$ \,& $T_c$ \,& $P_c$ \,& $\frac{P_c\, r_c}{T_c}$ \,& 
       $\mathcal{D}_1$ \,& $\mathcal{D}_3$ \,& $\mathcal{D}_5$
       \\
       \hline
       5\,&\,1 \,&\, 0.02 \,&\, 1.540814  \,&\, 0.161223\,&\, 0.0372253\,&\, 0.355764\,&\,  0.20474\,&\, 
       2.44204\,&\, -2.44204  \\
       6 \,&\, 1.1 \,&\, 0.05 \,&\, 1.28493  \,&\, 0.306192\,&\, 0.109039\,&\, 0.457581\,&\,  0.32596\,&\,
       1.5339\,&\, -1.5339 \\
       7 \,&\, 1.2 \,&\, 0.09 \,&\, 1.19821  \,&\, 0.456712\,&\,0.219385\,&\, 0.575571\,&\,  0.70392\,&\,
       0.7103\,&\, -0.7103\\
       8 \,&\, 1.3 \,&\, 0.5 \,&\, 1.13747  \,&\, 0.632947\,&\, 0.408496\,&\, 0.734109\,&\,3.7441 \,&\, 
       0.1335\,&\, -0.1335 \\
           \hline\hline
    \end{tabular}
     \caption{Numerical sets for critical physical quantities and coefficients $\mathcal{D}=1$ and $\mathcal{D}_i$ in $P - V$ critical behavior with $k =1$}
      \label{Tab3}
    \end{table*}

    \subsection{Gibbs free energy}
We now analyze our solution's global stability by investigating Gibbs free energy. Notably, this is a thermodynamic potential reckoned from the Euclidean action and an appropriate cut-off term. Using the Gibbs free energy sign, a comprehensive stability analysis can be done.  

In the extended phase space, $G=M-T\, S=H-T\, S=H-T \, S$ gives the Gibbs free energy. Practically speaking, it should be noted that any discontinuous behavior in the first- or second-order derivatives of the Gibbs energy leads to a first- or second-order phase transition in the system. Gibbs free energy is therefore specified by
\begin{widetext}
\begin{align}\label{r56}
G&=G(P, T)=H-TS=M -TS
   \nonumber
   \\
&\hspace{-4mm}=\frac{1}{16 (d-2) \Gamma \left(\frac{d+1}{2}\right) \sqrt{\gamma +Q^2 r_h^{2-2 d}}}\Biggl\{\pi ^{\frac{d-3}{2}} r_h^{-d-3} \Bigg(r_h^{2 d} \left((d-2) (d-1) k \sqrt{\gamma +Q^2 r_h^{2-2 d}}+2 r_h^2 \left(\gamma -8 \pi  P \sqrt{\gamma +Q^2 r_h^{2-2 d}}\right)\right)\nonumber\\[2mm]
&\hspace{-4mm}+2 \sqrt{\gamma } (d-2) Q
   r_h^{d+3} \sqrt{\frac{Q^2 r_h^{2-2 d}}{\gamma }+1} \text{ArcSinh}\left(\frac{Q r_h^{1-d}}{\sqrt{\gamma }}\right)+2 Q^2 r_h^4\Bigg)\Biggr\}\,.
\end{align}
\end{widetext}

\begin{figure*}[tbh!]
      	\centering{
       \includegraphics[scale=0.73]{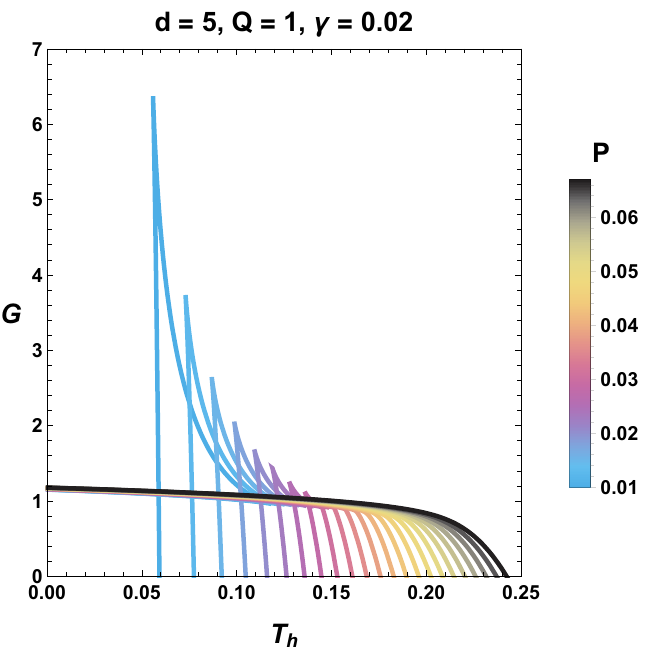} \hspace{2mm}
      	\includegraphics[scale=0.72]{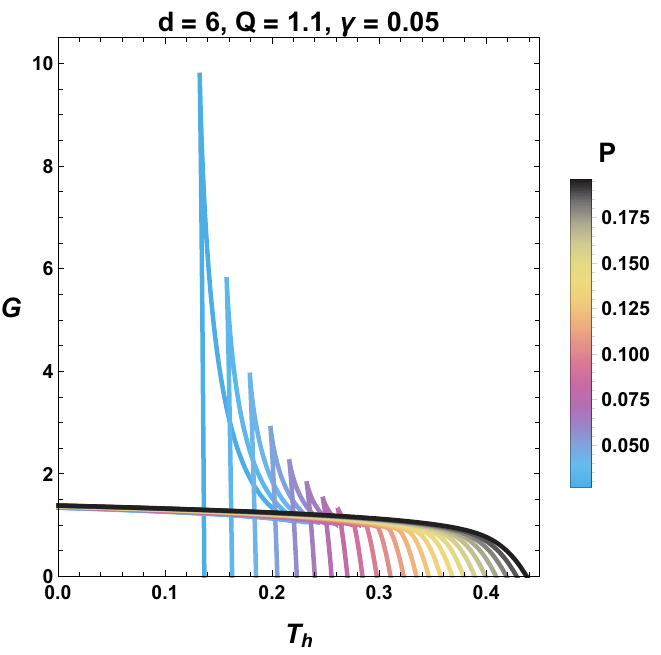} \hspace{2mm}
      }
       	\caption{$G-T$ diagram of the BH system for fixed values of the parameter space.}
      	\label{fig7}
      \end{figure*}
     
    The behavior of Gibbs free energy as a function of the horizon temperature is displayed in Fig.~$\ref{fig7}$. We can distinguish three regimes in compliance with the former $P-v$ analysis. In the case of $(P<P_c)$, a classic swallowtail phenomenon is instigated on the $G$-$T$ diagram, featuring a first-order small/large BH phase transition~\cite{Kubiznak:2012wp}.
    This means that the non-smooth points on the isobaric curves associated with the $G-T$ diagram are those of the extreme points joining the isobaric curve in the $T-r$ diagram. In the tracking phase of the scenario, the swallow's size gradually decreases and ultimately disappears as pressure increases. 
    Owing to the increase in pressure, the extreme points of $T$ on the isobaric curve move closer together to converge on the critical pressure $(P=P_c)$. As a result, any first-order phase transition in the system disappears completely. In the case of $P>P_c$, $G$ becomes a monotonic function of $T$, showing no phase transition in the system. The dimension of the system does not spoil such a global behavior.

\subsection{Critical exponents}
Critical exponents ideally predict the behavior of physical quantities in the neighborhood of the critical point. The computation of such exponents is of extreme interest, as they are independent of the physical system and can be treated as quasi-universal parameters. In this respect, attention is paid to the following terminology:
\begin{equation}
    t\equiv\frac{T}{T_c}-1,\quad\omega\equiv\frac{V}{V_c}-1,\quad p\equiv\frac{P}{P_c},
\end{equation}
where the critical thermodynamic volume $V_c$ is associated with the critical horizon radius $r_c$ by $V_c=(\omega_{d-2}/d-1)r_c^{d-1}$. In this way, the critical exponents are explicitly defined as follows~\cite{Kubiznak:2012wp}:
\begin{align}
    C_V&\propto\lvert t\rvert^{-\alpha}\,,\\[2mm]
    \eta&\propto\lvert t\rvert^{\lambda}\,,\\[2mm]
    \kappa_T&\propto\lvert t\rvert^{-\gamma}\,,\\[2mm]
    \lvert P-P_c\rvert&\propto\lvert V-V_c\rvert^\delta.\label{71}
    \end{align}

    Even at constant volume, the exponent $\alpha$ describes the behavior of specific heat $C_V= T\left(\frac{\partial S}{\partial T}\right)_V$. Since, from Eq.~\eqref{mass1}, entropy $S$ is independent of the Hawking temperature $T$, we conclude that exponent $\alpha = 0$.

     The exponent $\beta$ characterizes the behavior of the order parameter near the critical point. For the vdW system, this is defined as the difference $\eta$ of the volume of the gas phase and that of the liquid phase, i.e. $\eta=v_g-v_l\propto |t|^{\beta}$.
     In our language, this amounts to consider the difference between the volumes of the large and small BH phases. In order to estimate $\beta$, let us approximate the EoS near the critical point as
     \begin{equation}
         p=\mathcal{D}+\mathcal{D}_1\, t+\mathcal{D}_2\, \omega+\mathcal{D}_3\,t\omega+\mathcal{D}_4\,\omega^2+\mathcal{D}_5\,\omega^3+O(t\omega^2,\omega^4),
     \end{equation}
     where we have defined
     \begin{widetext}
           \begin{align}
         \mathcal{D}& \equiv\frac{1}{16 \pi  P_c r_c^2}\Biggl\{2 r_c \left(2 \pi  (d-2) T_c+r_c \sqrt{Q^2 r_c^{2-2 d}+\gamma}\right)+(3-d) (d-2) k\Biggr\},\\[2mm]
   \mathcal{D}_1& \equiv-\mathcal{D}_3=\frac{(d-2)}{4}\frac{ T_c}{ P_c r_c},\\[2mm]
   \mathcal{D}_2& \equiv\mathcal{D}_4 =0,\\[2mm]
   \mathcal{D}_5& \equiv \frac{1}{48 P_c r_c^2}\Biggl\{-\frac{(d-1) Q^2 r_c^4 \sqrt{Q^2 r_c^{2-2 d}+\gamma } \left(d (d+1) Q^4
   r_c^4+\gamma  (d+3) (2 d-1) Q^2 r_c^{2 d+2}+2 \gamma ^2 d (2 d-1) r_c^{4
   d}\right)}{\pi  \left(\gamma  r_c^{2 d}+Q^2 r_c^2\right){}^3}\nonumber\\[2mm]
   &-12 (d-2) r_c
   T_c+\frac{12 (d-3) (d-2) k}{\pi }\Biggr\}\,.
     \end{align}
     \end{widetext}
     Numerically, the dependencies of the coefficients $\mathcal{D}_i$ on the parameters $d$, $Q$ and $\gamma$ are presented in Tab.~$\ref{Tab3}$. 

     As the pressure remains constant during the phase transition, one has 
     \begin{equation}
         \mathcal{D}+\mathcal{D}_1 \,t+\mathcal{D}_3 \,t\omega_l+\mathcal{D}_5\,\omega_l^3=\mathcal{D}+\mathcal{D}_1\, t+\mathcal{D}_3 \,t\omega_s+\mathcal{D}_5\,\omega_s^3.\label{78}
     \end{equation}
     where $\omega_s$ and $\omega_l$ are the reduced volumes of the small and large BHs, respectively.

     In addition, Maxwell's equal area law is given by the following formula~\cite{Kubiznak:2012wp}
   \begin{equation}
       \int_{\omega_l}^{\omega_l}\omega\,\frac{\mathrm{d}p}{\mathrm{d}\omega}\,\mathrm{d}\omega=0\,,\label{79}
   \end{equation}
considering that 
\begin{equation}
    \frac{\mathrm{d}p}{\mathrm{d}\omega}=\mathcal{D}_3\, t+3\,\mathcal{D}_5\,\omega^2.\label{80}
\end{equation}
Exploiting Eqs. \eqref{79} and \eqref{80} yields the following equation:
\begin{equation}
    \mathcal{D}_3\, t(\omega_s^2-\omega_l^2)+ \frac{3}{2}\,\mathcal{D}_5(\omega_s^4-\omega_l^4)=0\,,
\end{equation}
from which it is possible to find an explicit link between $\omega_l$ and $\omega_s$ in the form
\begin{equation}
    \omega_l=-\omega_s=\sqrt{-\frac{\mathcal{D}_3}{\mathcal{D}_5}t}\,,\label{82}
\end{equation}
where the argument under the square root function remains positive. A quick look at Eq. \eqref{82}  yields the desired results, namely
\begin{equation}
    \eta=V _l-V_s=V_c(\omega_l-\omega_s)=2 V_c\,\omega_l\propto\sqrt{-t}\,,
\end{equation}
which provides $\beta=1/2$.

The exponent $\gamma$ describes the critical behavior of the isothermal compressibility $\kappa_T$ given explicitly by
\begin{equation}
    \kappa_T=-\frac{1}{V}\frac{\partial V}{\partial P}\biggr\rvert_{V_c}=|t|^{-\gamma}.
    \end{equation}
In the present case, simple calculations allow us to show   
    \begin{equation}
    \kappa_T-\frac{1}{P_c}\frac{1}{\frac{\partial p}{\partial \omega}}\biggr\rvert_{\omega =0}\propto\, \frac{2r_c}{T_c}t^{-1}\,,
\end{equation}
giving rise to $\gamma=1$. 

Finally, the exponent $\delta$ describes the critical behavior of Eq. \eqref{71} on the critical isotherm $T=T_c$. So, the shape of the critical isotherm is defined at $t=0$, providing the following finding:
\begin{equation}
    \lvert P-P_c\rvert= P_c  \lvert p-1\rvert= P_c  \,\lvert \mathcal{D}_5\, \omega^3\rvert=\frac{P_c\,\lvert \mathcal{D}_5\rvert}{V_c^3} \lvert V-V_c\rvert^3\,,
\end{equation}
which easily proves $\delta=3$.

Based on the above results, we conclude that 
these four critical exponents take the same values as those previously obtained for charged AdS BHs~\cite{Kubiznak:2012wp}. This, in turn, reveals that the CDF does not influence the critical exponents in the same way as the quintessential dark energy. So both the universality profile of VdW-type phase transitions and the values of the critical exponents for AdS BHs have been verified.

\subsection{Sparsity of BH radiation}

As a further property of BHs and Hawking radiation flow, we move on to the computation of sparsity, 
defined as the average time gap between the emission of successive quanta. Compared to the black body, Hawking radiation appears significantly more sparse, as shown, e.g., in~\cite{Spars}. This is a key feature that distinguishes between the two systems classes. 

For $d$-dimensional BHs, sparsity is quantified by the parameter,
\begin{equation}
\label{sparseq}
    \eta =\frac{C}{\Tilde{g} }\left(\frac{\lambda_t^{d-2} }{\mathcal{A}_{eff}}\right),
\end{equation}
which reduces to the well-known definition for $d=4$~\cite{Spars}. Here, $C$ is a dimensionless constant, $\Tilde{g}$ the spin degeneracy factor of the emitted quanta
$\lambda_t=2\pi/T$ their the thermal
wavelength and $A_{eff}=27A/4$ the effective BH area. For the simplest case of $(1+3)$-dimensional Schwarzschild BHs, assuming the emission of massless bosons, one finds the constant value $\eta=64\pi^3/27\simeq73.49$. For comparison, we remind that $\eta\ll1$ for black bodies.  

Corrections induced on Eq.~\eqref{sparseq} by generalized entropies and/or uncertainty relations have been studied in~\cite{Cor1,Cor2,Cor3,Luciano:2023fyr,Luciano23}. Moreover, the computation of sparsity in generic $D+1$-dimensional Tangherlini BHs has been developed in~\cite{SpHiDim}. The purpose here is to 
study how Eq.~\eqref{sparseq} appears for charged AdS BHs with a surrounding MCG in higher dimensions. Toward this end, we observe that direct substitution of the modified Hawking temperature in Sec.~\ref{secther} into Eq.~\eqref{sparseq} gives
\begin{widetext}
    \begin{equation}
      \eta = \frac{1}{27} 2^{3 d-5} \pi ^{\frac{1}{2} (3 d-7)}\mathcal{A}_1 \left(\frac{d-2}{2^{\frac{1}{2-d}} (d-3) (d-2) k \left(\mathcal{A}_2\right)^{\frac{1}{2-d}}-2^{\frac{1}{d-2}+1} \left(\mathcal{A}_2\right)^{\frac{1}{d-2}} \left(\sqrt{\gamma +Q^2
   \left(2^{\frac{1}{d-2}} \left(\mathcal{A}_2\right)^{\frac{1}{d-2}}\right)^{2-2 d}}+\Lambda \right)}\right)^{d-2}\,,
    \end{equation}
\end{widetext}
where we have defined
\begin{align}
   \mathcal{A}_1&= \Gamma \left(\frac{d-1}{2}\right) \left(2^{\frac{1}{d-2}} \mathcal{A}_2^{\frac{1}{d-2}}\right)^{2-d}\,,\\
  \mathcal{A}_2 &= \pi ^{\frac{1}{2}-\frac{d}{2}}\, S \,\Gamma
   \left(\frac{d-1}{2}\right)\,.
\end{align}
\begin{figure}[t]
      	\centering{
      	\includegraphics[scale=0.8]{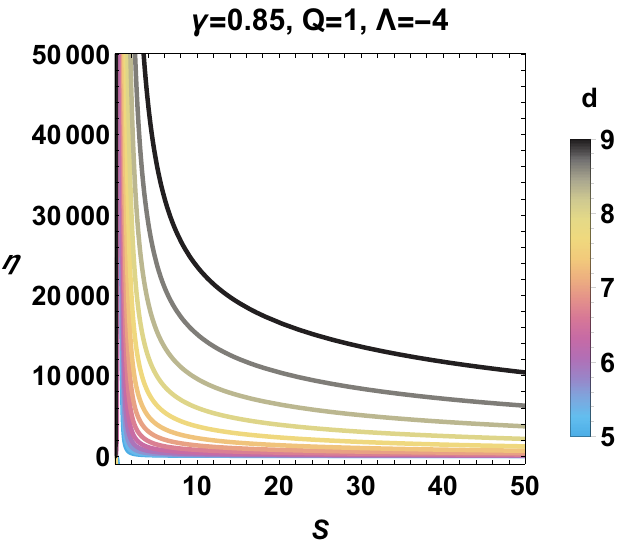} 
      	}
      	\caption{Plot of sparsity $\eta$ against $S$ for various values of the parameter $d$.}
      	\label{spars}
      \end{figure}

We can gain some interesting insights by looking at the plot in Fig.~\ref{spars}.
Unlike Schwarzschild BHs, sparsity varies (in particular decreases) for increasing $S$, that is, for increasing BH size. For $S$ large enough, we have $\eta\ll1$: in this regime, the radiation emitted by AdS BHs with surrounding MCG becomes almost classical
and fully comparable to a black body spectrum. On the other hand, we find that, for a given $S$, the higher the dimension of the BH, the more sparse the radiation, and vice-versa. This is due to the presence of the surrounding MCG, which contributes to slowing down the emission of quanta and consequently enhances the radiation's sparsity as $d$ increases. For comparison, we notice that in~\cite{SpHiDim} - where no Chaplygin gas is considered - sparsity is lost in high dimensions. Thus, such a result provides an original and peculiar feature of the present analysis. 

\section{Thermal geometries}
\label{sec5}
Following the facts outlined in the Introduction, we now study phase transitions from a geometrothermodynamic perspective. Based on the specific scalar curvature adopted, GT tools are employed to predict physical limitation and/or divergent points that belong to the set of roots $C_{P, Q}= 0$ or $C_{P, Q}^{-1}= 0$, respectively. To simplify the applied GT algorithm, it is useful to consider the space of physical limitation and divergent points as a moduli space spanned by $\lbrace r^\star_i, r_i^ {\text{div}}\rbrace$, respectively.  
The shape of the metric is examined with a special focus on the Weinhold, Ruppeiner, HPEM, and Quevedo models. 

\subsection{Weinhold and Ruppeiner formalisms}
In the mass representation, Weinhold shape takes the form~\cite{Weinhold1,Weinhold2,Soroushfar:2016nbu}
\begin{equation}
    g_{jk}^W=\partial_j\partial_k\, M(S, Q, \ell)\,.
\end{equation}
For a charged AdS BH, the line element appears as follows:
\begin{align}
\mathrm{d}s_W^2&=M_{SS}\,\mathrm{d}S^2+M_{\ell\ell}\,\mathrm{d}\ell^2+M_{QQ}\,\mathrm{d}Q^2
+2M_{S\ell}\,\mathrm{d}S\mathrm{d}\ell\nonumber\\
&+2M_{SQ}\, \mathrm{d}S\mathrm{d}Q+2M_{\ell Q}\,\mathrm{d}\ell\mathrm{d}Q\,,
\end{align}
In terms of the mass matrix representation, the formulation is given explicitly as
\begin{equation}
    g^W= \begin{pmatrix}
M_{SS} & M_{S\ell} & M_{SQ} \\
M_{\ell S} & M_{\ell\ell} & 0 \\
M_{Q S} & 0 & M_{Q Q}\,.
\end{pmatrix}
\end{equation}

Similarly, in Ruppeiner formalism, one considers entropy as basic thermodynamic potential, i.e.~\cite{Rupp1,Rupp2}
\begin{equation}
    g_{jk}^R=\partial_j\partial_k\, S.
\end{equation}
It is worth noting that the Ruppeiner metric is linked to the Weinhold metric by a conformal transformation, yielding the following defining expression~\cite{Mrugala1984OnEO}:
\begin{equation}
    \mathrm{d}s_R^2=\frac{1}{T}\mathrm{d}s_W^2
\end{equation}
Therefore, in terms of mass matrix representation, one has 
\begin{equation}
    g^R=\frac{1}{T} \begin{pmatrix}
M_{SS} & M_{S\ell} & M_{SQ} \\
M_{\ell S} & M_{\ell\ell} & 0 \\
M_{Q S} & 0 & M_{Q Q}
\end{pmatrix}.
\end{equation}

We emphasize that our next treatment will be carried out in terms of pressure. Using Eq.~\eqref{pl}, we can relate differentiation with respect to AdS length and pressure by the following expression:
\begin{eqnarray}
    \partial\ell=-\frac{(d-1) (d-2)}{32\pi\ell}\frac{\partial P}{P^2}.
\end{eqnarray}
 \begin{figure*}[t!]
    \centering
    \begin{subfigure}[b]{0.5\textwidth}
        \centering
        \includegraphics[scale=0.77]{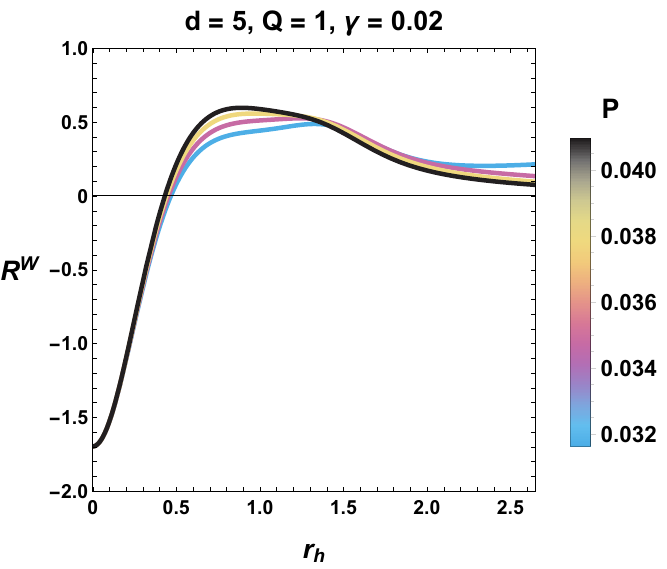}
        \caption{Physical limitation point}
        \label{fig8a}
    \end{subfigure}%
    \hfill
    \begin{subfigure}[b]{0.5\textwidth}
        \centering
        \includegraphics[scale=0.75]{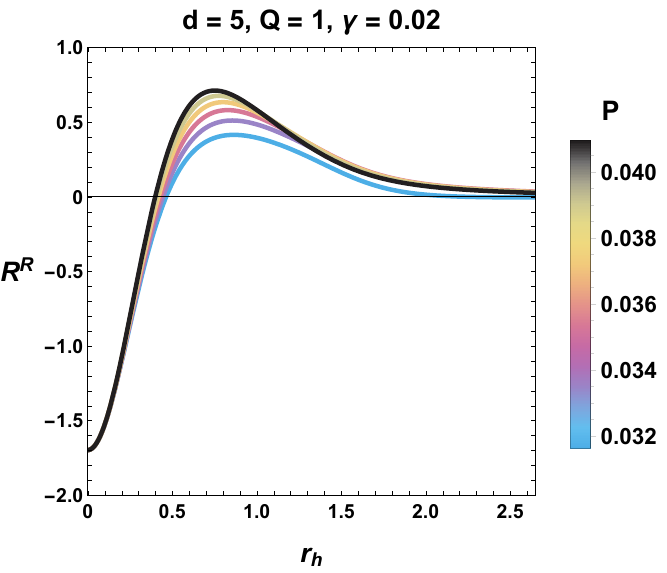}
        \caption{Physical limitation point}
        \label{fig8b}
    \end{subfigure}%
    \caption{Curvature scalar variation of Weinhold (left panel) and Ruppeiner (right panel) metrics for $0.8P_c\leq P\leq1.2 P_c$}
    \label{fig8}
\end{figure*}
As stressed above, to conveniently reveal the physical limitation and critical phase transition points of the heat capacity from the GT perspective, it is necessary to consider the set of roots of the equations $numer(R^i)=0$ or $denom(R^i)=0$, respectively, where $i$ labels the different types of GTs applied tools. 
The explicit expressions of Weinhold and Ruppeiner curvatures are awkward to exhibit and add little to the plots in Fig.~\ref{fig8}, which catch all the interesting features by themselves. From these diagrams, we see that the use of Weinhold and Ruppeiner tools in the present study successfully predicts the physical limitation point of BHs, namely $0.549601\leq r^\star\leq0.55514$ within the interval $0.0316415\leq P\leq0.0409478$. 
This shows that the physical limitation point is always present, as observed in the heat capacity analysis. 

We can also infer direct information about the character of the interaction among the micro-constituents of BHs looking at the sign of the scalar curvature. For $r$ above the physical limitation point, both Weinhold and Ruppeiner curvatures take positive values. 
According to~\cite{1999JPhA,Wei:2015iwa}, the repulsive interaction dominates the thermodynamic system. As the micro-molecules move apart due to repulsion and the size of the system increases, the interaction becomes weaker. For $r$ large enough, the scalar curvature $R\rightarrow0$, and the BH behave like the classical ideal gas. 

\subsection{HPEM and Quevedo tools}
Despite these virtues, the usage of Weinhold and Ruppeiner formalisms does not allow the identification of the critical phase transition points analyzed for the heat capacity phase transition. Moreover, the phase space and the metric structure analysis of~\cite{salamon1983group,PhysRevA.41.3156} suggests that these two metrics are not invariant under Legendre transformations.       

It might be useful to apply GT tools to get a better approach to the BH phase transition analysis. For instance, in~\cite{Quevedo:2006xk}, Quevedo attempted to unify the geometric properties of the phase space and the space of equilibrium states (Quevedo case I and II). Furthermore, Hendi–Panahiyan–Eslam–Momennia (HPEM) introduced another kind of metric, which builds a geometrical phase space by thermodynamic quantities~\cite{Hendi:2015rja, Hendi:2015fya, Hendi:2015xya, Hendi:2015hoa, Hendi:2015pda}. 

The Quevedo metric is expressed as follows
\begin{equation}
    g=\left( E^c\frac{\partial\Phi}{\partial E^c}\right)\left(\eta_{ab}\delta^{bc}\frac{\partial ^2\Phi}{\partial E^c\partial E^d}\mathrm{d}E^a\,\mathrm{d}E^d\right)\,,
\end{equation}
where
\begin{equation}
    \frac{\partial \Phi}{\partial E^c}=\delta_{cb}\, I^b.
\end{equation}
Within this structure, $\Phi$, $I^b$, and $E^a$ represent the thermodynamic potential, the intensive and extensive variables, respectively, On the other hand, the generalized HPEM metric with $n$ extensive variables ($n\ge2$) is given in such a way as~\cite{Mrugala1984OnEO, Hendi:2015rja, Hendi:2015fya, Hendi:2015xya, Hendi:2015hoa, Hendi:2015pda}
\begin{equation}
    \mathrm{d}S^2_{HPEM}=\frac{S\,M_S}{\left(\prod_{i=2}^n\frac{\partial^2M}{\partial \xi_i^2}\right)^3}\left(-M_{SS}\mathrm{d}S^2+\sum_{i=2}^n\left(\frac{\partial^2M}{\partial \xi_i^2}\right)\mathrm{d}\xi_i^2\right).
\end{equation}
Here, $\xi_i(\xi_i\neq S)$, $M_S=\frac{\partial M}{\partial S}$, and $M_{SS}=\frac{\partial^2 M}{\partial S^2}$ are extensive parameters. 

Following on from the above considerations, the background of HPEM and Quevedo metrics can be written collectively  as follows~\cite{Mrugala1984OnEO,Hendi:2015rja, Hendi:2015fya, Hendi:2015xya, Hendi:2015hoa, Hendi:2015pda,Quevedo:2006xk}:
\begin{widetext}
    \begin{align}
    \mathrm{d}S^2_{HPEM}&=\frac{S\,M_S}{\left(\frac{\partial^2M}{\partial \ell^2}\frac{\partial^2M}{\partial Q^2}\right)^3}\left(-M_{SS}\mathrm{d}S^2+M_{\ell\ell}\mathrm{d}\ell^2+M_{QQ}\mathrm{d}Q^2\right),\\[2mm]
    \mathrm{d}S^2_{QI}&=\left(S M_S+\ell M_{\ell}+Q M_{Q} \right)\left(-M_{SS}\mathrm{d}S^2+M_{\ell\ell}\mathrm{d}\ell^2+M_{QQ}\mathrm{d}Q^2\right),\\[2mm]
    \mathrm{d}S^2_{QII}&=S M_S\left(-M_{SS}\mathrm{d}S^2+M_{\ell\ell}\mathrm{d}\ell^2+M_{QQ}\mathrm{d}Q^2\right).
\end{align}
\end{widetext}

 \begin{figure*}[t!]
    \centering
    \begin{subfigure}[b]{0.5\textwidth}
        \centering
        \includegraphics[scale=0.8]{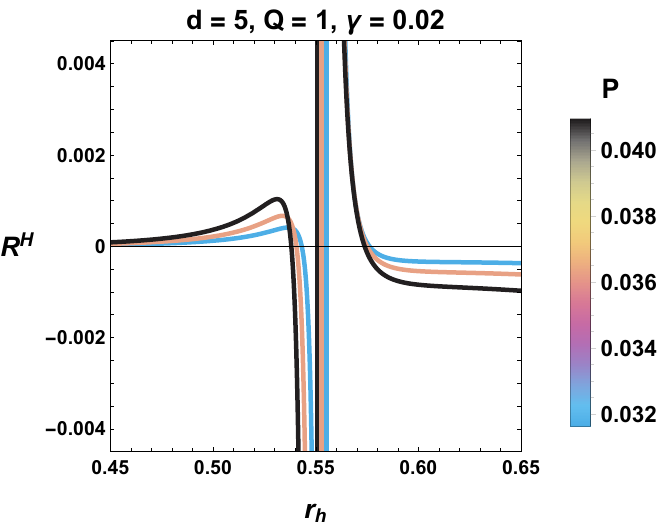}
        \caption{Physical limitation point}
        \label{fig9a}
    \end{subfigure}%
    \hfill
    \begin{subfigure}[b]{0.5\textwidth}
        \centering
        \includegraphics[scale=0.75]{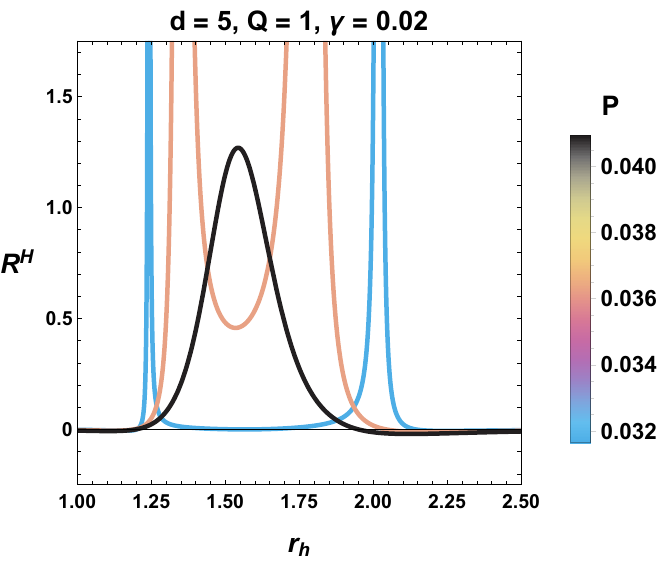}
        \caption{phase transition critical point}
        \label{fig9b}
    \end{subfigure}%
    \caption{Variation of the curvature scalar of the HPEM metric against $r_h$ for the interval value $0.8P_c\leq P\leq1.2 P_c$.}
    \label{fig9}
\end{figure*}
      \begin{figure*}[t!]
    \centering
    \begin{subfigure}[b]{0.5\textwidth}
        \centering
        \includegraphics[scale=0.77]{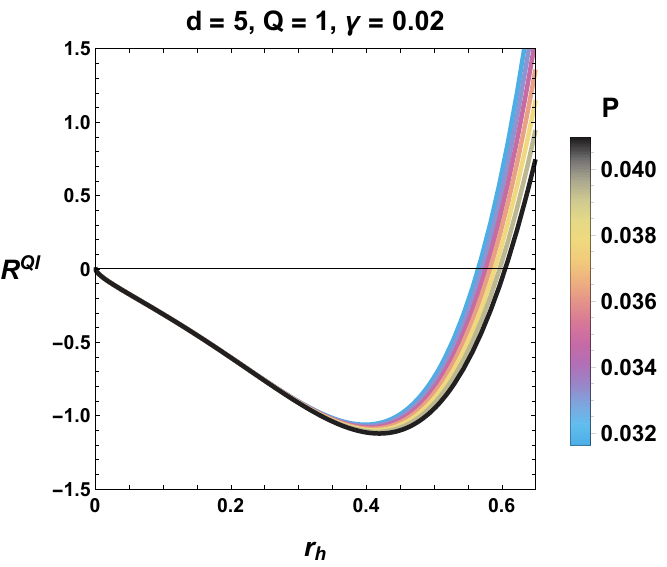}
        \caption{Physical limitation point}
        \label{fig10a}
    \end{subfigure}%
    \hfill
    \begin{subfigure}[b]{0.5\textwidth}
        \centering
        \includegraphics[scale=0.79]{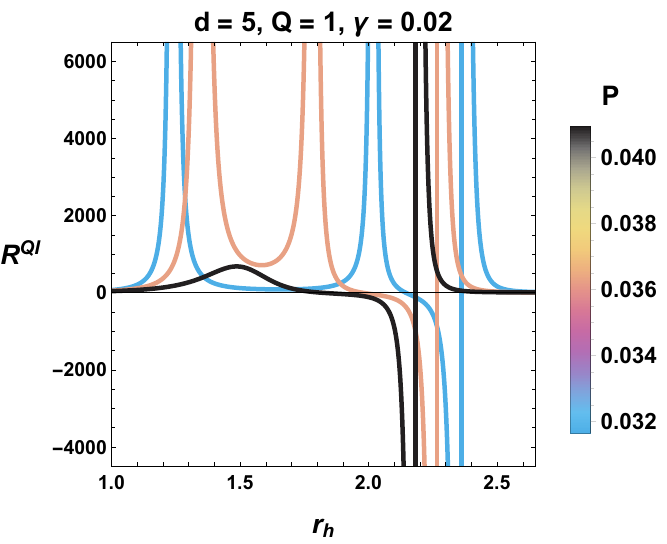}
        \caption{phase transition critical point}
        \label{fig10b}
    \end{subfigure}%
    \caption{Variation of the curvature scalar of the Quevedo class I metric against $r_h$ for the interval value $0.8P_c\leq P\leq1.2 P_c$.}
    \label{fig10}
\end{figure*} 
 \begin{figure*}[t!]
    \centering
    \begin{subfigure}[b]{0.5\textwidth}
        \centering
        \includegraphics[scale=0.77]{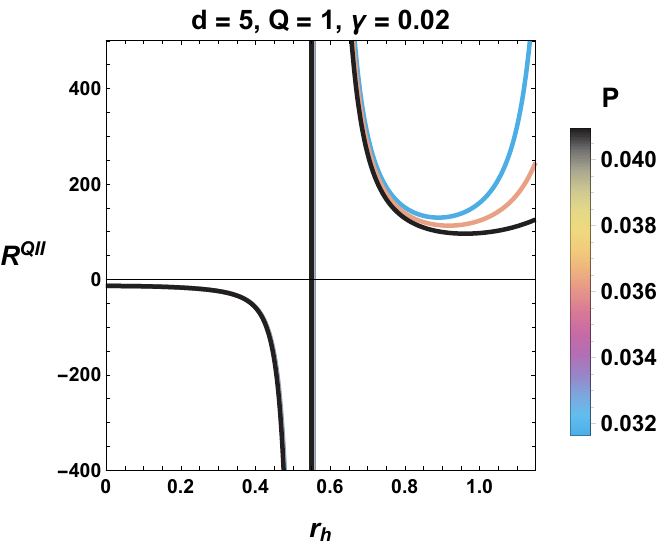}
        \caption{Physical limitation point}
        \label{fig11a}
    \end{subfigure}%
    \hfill
    \begin{subfigure}[b]{0.5\textwidth}
        \centering
        \includegraphics[scale=0.79]{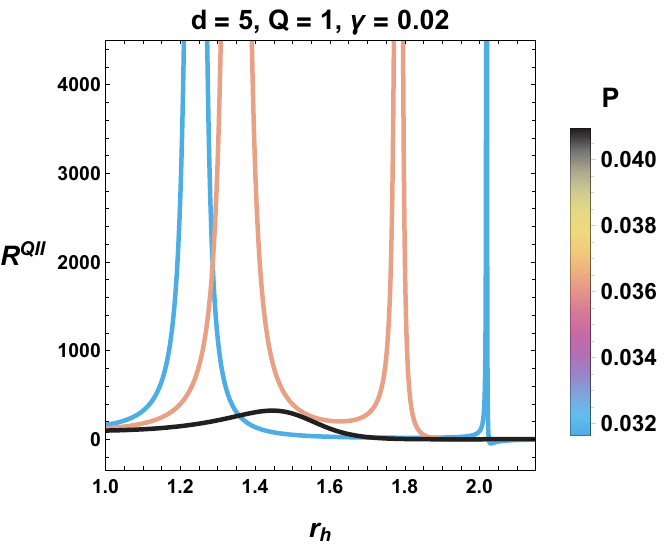}
        \caption{phase transition critical point}
        \label{fig11b}
    \end{subfigure}%
    \caption{Variation of the curvature scalar of the Quevedo class II metric against $r_h$ for the interval value $0.8P_c\leq P\leq1.2 P_c$.}
    \label{fig11}
\end{figure*} 
In turn, the corresponding Ricci scalars  have the following denominator~\cite{Mrugala1984OnEO,Hendi:2015rja, Hendi:2015fya, Hendi:2015xya, Hendi:2015hoa, Hendi:2015pda}
\begin{widetext}
    \begin{equation}
   \text{denom}(R) = \begin{cases}
     2M_{SS}^2 S^3 M_{S}^3\hspace{4.62cm} \text{HPEM}  \,, \\[2mm]
      2M_{SS}^2M_{QQ}^2M_{\ell\ell}^2\left(S M_S+Q M_Q+ \ell M_{\ell}\right)^3 \quad \text{Quevedo Case I}\,,\\[2mm]
      2S^3 M_{SS}^2M_{QQ}^2M_{\ell\ell}^2 M_S^3 \hspace{3.15cm}\text{Quevedo Case II}\,.
    \end{cases}\,.
\end{equation}
\end{widetext}
      
HPEM geometry detects all the specific points corresponding to the heat capacity phase transition (see Fig.~\ref{fig9}). In particular, the physical limitation point appears in the realm of this geometry as a root of the scalar curvature inverse ($\text{denom}(R^{HPEM})=0$). The physical limitation point is always present, as proven previously. The HPEM scalar curvature also involves two positive divergent points for $P<P_c$ that perfectly match those of the heat capacity. As $P$ increases, such points approach each other until they coincide at  $P=P_c$. Finally, 
the case $P>P_c$ generates the absence of any divergent point across a finite curve of the HPEM scalar curvature. 

On the same footing, using Quevedo class $I$ allows us to reveal both the physical limitation and critical phase transition points (see Fig.~\ref{fig10}). In particular, the physical limitation point is detected as a set of roots of Quevedo class I scalar curvature $(\text{numer}(R^{QI})=0)$ and is everywhere present. This shows a one-to-one correspondence between the physical limitation point of the heat capacity and that explored within the set of roots $\text{numer}(R^{QI})=0$. 
In addition, Quevedo class I involves three divergent points below the critical pressure, two of which are positive. These points are the same as the roots of $1/C_{P, Q}=0$. The scenario that emerges here is that, as the pressure increases to the critical value, the two positive divergent points close on each other until the divergence behavior disappears, producing a convergent behavior for the scalar curvature in the case $P>P_c$, where no second-order phase transition may take place.

The last tool concerns Quevedo class II, which can capture physical limitations and phase transition critical points (see Fig.~\ref{fig11}). 
Exploring physical limitation emerges with the smaller singular point of the Ricci scalar regarding Quevedo class II. In contrast, transition critical points are associated with positive divergent points of scalar curvature of Quevedo class II. As before, such points tend to collapse as $P\rightarrow P_c$, resulting in the disappearance of any divergence in the case $P>P_c$. 

Therefore, it can be claimed that one can extract more information from HPEM and Quevedo cases than from Weinhold and Ruppeiner geometries.

\section{Conclusion}
\label{conc}
A thorough grasp of dark fluids with cosmological models in terms of dark energy and dark matter governs the description of the Universe in a phenomenological way, providing a meaningful connection with string theory. In the realm of CDF structure with Chaplygin-like EoS $p=-\frac{
\gamma}{\rho}$, we derived a new higher-dimensional, static, and spherically symmetric AdS BH solution. The explored BH solution has offered a way to inspect the satisfaction and validation of classical energy conditions such as WEC, NEC, and SEC. The findings show that the WEC and NEC are indeed satisfied, while violations of the SEC are noted. In parallel, the analysis of curvature singularities tools, namely Ricci and Kretschmann scalars, has been applied, demonstrating the uniqueness and physical validity of the BH solution.

On the side of thermodynamic features, the Smarr relation has been inspected for higher dimensions in the extended phase space endowed with the standard definition of pressure in terms of cosmological constant. In this framework, the $P$-$V$ behavior has been examined. It has been proven that our solution still exhibits a small/large BH phase transition, analogous to the vdW liquid/gas change of state. Further study in this direction has been conducted by computing the critical exponents. 

As a final step, GT tools have been employed to confirm our thermodynamic findings independently. Specifically, we have applied Weinhlold, Ruppeiner, HPEM, and Quevedo classes I and II. Our analysis has revealed consistent results between the GT scalar curvature and the physical limitation (zero) point and/or the critical phase transition (divergent) points of the heat capacity. Valuable insights into the behavior of micro interactions have been obtained by looking at the sign of the metric curvature. 

This work prompts several inquiries and presents an opportunity to delve into various physical aspects, especially considering the novel BH solution formulated within higher-dimensional frameworks. Exploring concepts like thermodynamic topology and Joule-Thomson expansion could yield valuable insights. Additionally, investigating topics such as quasinormal ringing and optical properties holds promise for future research endeavors.
Work is progressing in these directions and will be presented elsewhere.

\section*{Acknowledgments}
The authors are grateful to the anonymous Referees, who contributed to improving the quality of the manuscript with their comments and recommendations. GGL acknowledges the Spanish "Ministerio de Universidades" for the Maria Zambrano fellowship awarded and funding received from the European Union—NextGenerationEU. He also acknowledges participation in LISA. J.R. acknowledges Grant No. FA-F-2021-510 of the Uzbekistan Agency for Innovative Development and Silesian University in Opava for hospitality.

\bibliographystyle{apsrev4-2}
\bibliography{Library}
\end{document}